\documentclass[manuscript]{aastex}
\usepackage{graphicx}

\newcommand{\rsun}{R_{\Sun}}
\newcommand{\mydeg}{^{\circ}}

\shorttitle{ForeCAT CME Deflection}
\shortauthors{Kay, Opher and Evans}

\begin{document}


\title{Forecasting a Coronal Mass Ejection's Altered Trajectory: ForeCAT}


\author{C. Kay and M. Opher}
\affil{Astronomy Department, Boston University,
    Boston, MA 02215}
\email{ckay@bu.edu}

\author{R. M. Evans}
\affil{NASA Goddard Space Flight Center, Space Weather Lab, Greenbelt, MD 20771}




\begin{abstract}
To predict whether a coronal mass ejection (CME) will impact Earth, the effects of the background on the CME's trajectory must be taken into account. We develop a model, ForeCAT (Forecasting a CME's Altered Trajectory), of CME deflection due to magnetic forces.  ForeCAT includes CME expansion, a three-part propagation model, and the effects of drag on the CME's deflection. Given the background solar wind conditions, the launch site of the CME, and the properties of the CME (mass, final propagation speed, initial radius, and initial magnetic strength), ForeCAT predicts the deflection of the CME.  Two different magnetic backgrounds are considered: a scaled background based on type II radio burst profiles and a Potential Field Source Surface (PFSS) background.  For a scaled background where the CME is launched from an active region located between a CH and streamer region the strong magnetic gradients cause a deflection of 8.1$\mydeg$ in latitude and 26.4$\mydeg$ in longitude for a 10$^{15}$ g CME propagating out to 1 AU. Using the PFSS background, which captures the variation of the streamer belt position with height, leads to a deflection of 1.6$\mydeg$ in latitude and 4.1$\mydeg$ in longitude for the control case.  Varying the CME's input parameters within observed ranges leads to the majority of CMEs reaching the streamer belt within the first few solar radii. For these specific backgrounds, the streamer belt acts like a potential well that forces the CME into an equilibrium angular position.

\end{abstract}




\section{Introduction}
The Sun explosively releases magnetized plasma known as coronal mass ejections (CMEs).  \citet{Gop09a} cataloged CMEs observed with the Solar and Heliospheric Observatory (SOHO) mission's Large Angle and Spectrometric Coronagraph (LASCO).  This catalog includes CMEs with a wide range of speeds (200-2500 km s$^{-1}$), masses (10$^{13}$-10$^{16}$ g) and kinetic energies (10$^{27}$-10$^{33}$ ergs). Earth-directed CMEs can drive space weather phenomena, producing aurora but also potentially damaging power grids.  Shocks driven by CMEs can accelerate particles.  At Earth these energetic particles can damage satellites and harm astronauts.  The better we understand the trajectory of a CME through the heliosphere, the better we can predict the effects at Earth and throughout the rest of the heliosphere. 

Since the beginning of CME observations in the 1970s, CME deflections have been observed \citep{Mac86}.  \citet{CaB04} and \citet{Kil09} discuss the trend of high latitude CMEs deflecting toward the equator during solar minimum conditions. Both attribute the deflection to polar coronal holes (CHs).  \citet{CaB04} emphasize the role of the fast wind affecting the CME's expansion. \citet{Kil09} suggest that CMEs cannot penetrate the polar CH magnetic fields which then guide the CME to the equator.  \citet{Kil09} also note the correlation between the direction of CME deflections and the decreased tilt of the heliospheric current sheet (HCS) at solar minimum.  At other times of the the solar cycle the increased complexity of the HCS configuration may lead to more variation in the direction of deflection.  \citet{Xie09} find that slow CMEs ($\le$ 400 km s$^{-1}$) follow a pattern of deflection toward the streamer belt (SB) but some fast CMEs move away from the SB.  \citet{Xie09} observe a correlation between the deflection of the CME and the distance between the CME source and the SB for the slow CMEs, but find no such correlation for the fast CMEs.  They also find that fast CMEs statistically tend to deflect less than slow ones. 

Recent observational studies show that CMEs can undergo strong deflections close to the Sun, however, below 5 $\rsun$ deflection cannot be distinguished from nonuniform expansion. Longitudinal deflections are observable using multiple coronagraph viewpoints after the launch of the Solar TErrestrial RElations Observatory (STEREO) with the Sun Earth Connection Coronal and Heliospheric Investigation (SECCHI). These observations confirmed that deflections can occur in both longitude and latitude \citep{Isa13,Liu10a,Liu10b,Lug10,Rod11}.  \citet{Byr10} reconstruct the 2008 December 12 CME in three dimensions (3D) using an elliptical tie-pointing method.  By matching the positions of edges in STEREO Ahead and Behind images they fit a 3D ellipsoid to the CME.  They estimate a latitudinal change of 30$\mydeg$ in the midpoint of their CME front during propagation up to 7 $\rsun$, but beyond this distance the latitude remains relatively constant. \citet{Liu10b} reconstruct the 3D behavior of several events using geometric triangulation: the forward modeling of a flux rope-like structure with self-similar expansion \citep{The06,The09}.  \citet{Liu10b} find a 13$\mydeg$ westward deflection within 15 $\rsun$ for the 2007 November 14 CME and about 10$\mydeg$ westward within 20 $\rsun$ for the 2008 December 12 CME, but do not address the latitudinal deflection calculated by \citet{Byr10}.  \citet{Liu10b} suggest that this systematic westward deflection may be a universal feature due to the magnetic field connecting the Sun and the CME.  If the solar magnetic field is frozen into the plasma of the CME, then corotation with the Sun would explain this motion.  The east-west asymmetry driven by a systematic westward deflection of CMEs was first observed in \citet{Wan04}.  \citet{Wan04} attribute the deflection to the Parker spiral and the speed of CMEs: CMEs traveling faster than the solar wind will deflect to the east and CMEs traveling slower than the solar wind will deflect to the west.  \citet{Isa13} use a combination of forward modeling of STEREO-SECCHI and SOHO-LASCO coronagraph images and Grad-Shafranov reconstruction to reconstruct the full three dimensional trajectory of a CME out to 1 AU.  \citet{Isa13} reconstruct 15 CMEs from between the minimum of Solar Cycle 23 and the maximum of Solar Cycle 24.  The latitudinal deflection of these CME far exceeds the longitudinal deflections.  Latitudinal deflections up to 35$\mydeg$ are observed and the maximum longitudinal deflection is only 5.4$\mydeg$.

CME deflection is also studied through the use of magnetohydrodynamic (MHD) simulations.  \citet{Lug11} present a MHD simulation of the 2005 August 22 CME using the Space Weather Modeling Framework \citep{Tot11}.  This CME was launched from an anemone active region (AR) within a CH.  \citet{Lug11} find that magnetic forces drive a deflection of 10-15$\mydeg$ within 8 $\rsun$ which is smaller than the 40-50$\mydeg$ expected from observations.  The simulated CME is initiated with an out-of-equilibrium flux rope so that the CME does not match the observations within three solar radii of the Sun.  The simulated CME reaches its maximum speed of 1500 km s$^{-1}$, only 1.5 minutes after initiation, but beyond 3 $\rsun$ the propagation speed matches the observed value of 1250 km s$^{-1}$. \citet{Lug11} note that the difference in propagation at low heights, where the magnetic forces should be the strongest, could explain some of the discrepancy between the observed and simulated deflections.  \citet{Zuc12} compare a MHD simulation to the 2009 September 21 CME which was observed to deflect 15$\mydeg$ toward the HCS.  In their MHD simulation, reconnection creates an imbalance in the magnetic pressure and tension forces causing the CME to deflect toward the SB.

A CH's influence on a CME has been quantified by defining force vectors based on the CH parameters.  To study correlations between a CME's deflection and the distance, $r$, from its source location to a CH with area, $A$, \citet{Cre06} introduce a force, $\textbf{F}=A/r \; \hat{r}$. This force points in $\hat{r}$, defined as the direction pointing away from the CH toward the CME.  They find a correlation between the direction of \textbf{F} and the direction of the CME deflection suggesting that CHs do influence the CME motion.  \citet{Gop09b} define an ``coronal hole influence parameter'' (CHIP) similar to the force of \citet{Cre06}, but also incorporate the magnetic field strength (B) of the CH, $\textbf{F}=B^2A / r \; \hat{r}$.   \citet{Gop09b} find good agreement between the direction a CME propagates after it deflects and the direction given by vector sum of all the individual F-vectors from nearby CHs. \citet{Gop10} update the r-dependence to $r^{-2}$ giving a final form $\textbf{F}=B^2 A /r^2 \; \hat{r}$.  \citet{Moh12} compare Solar Cycle 23 CMEs originating from disk center with their CHIP parameter as a function of the solar cycle and CME type.  Driverless shocks tend to have the largest CHIP value, magnetic clouds (MC) the smallest, with non-MCs falling in between.  The CHIP values are smallest during the rising phase.  \citet{Moh12} suggest CHs deflect CMEs away from the Sun-Earth line which provides support for the idea that all CMEs may be flux ropes, the distinction between MCs, non-MCs and driverless shocks being a matter of viewing perspective. 

\citet{She11} and \citet{Gui11} consider gradients in the magnetic energy density of the background solar corona as an explanation for the observed deflection.  At the distances of their observations ($\ge 1.5 \rsun$) these gradients point toward the streamer region.  During solar minimum conditions the streamer region is generally centered near the equator so mainly latitudinal deflections will occur.  At other times, the coronal magnetic field becomes more complex so a wider variety of gradient directions exists.  \citet{She11} present a theoretical approach that compares favorably with observations. \citet{Gui11} extend the work with additional observations and find that the direction of deflection tends to agree with the direction of the background gradients. 

Similar to the deflection of CMEs, \citet{Pan11} investigate the rolling motion of prominences/filaments.  They find that the prominences tend to roll away from CHs before they form flux ropes.  \citet{Pan11} suggest that the filament motion could be explained by local magnetic force imbalances within the filament arcade, whereas the non-radial motion of CMEs would result from similar imbalances on global scales.

This paper presents a model for CME deflection near the Sun by considering the effects of magnetic pressure gradients as well as magnetic tension.  Magnetic energy dominates the free energy budget of the ambient plasma in the lower corona, so magnetic forces play an important role in the deflection of CMEs near the Sun.  The closer to the Sun a CME is, the stronger the surrounding coronal magnetic fields and therefore the stronger the forces that act upon a CME. The magnetic field strength falls off quickly with distance so the magnetic forces should as well.  Other effects, not included in ForeCAT, can cause deflection such as interactions with other CMEs propagating through the interplanetary medium  \citep{Lug12, Tem12} or variations in the speed of the background solar wind.  Spatial velocity variations can distort the shape of a CME, seen in observations \citep{Sav10} and in MHD models \citep{Wan03}.  If unbalanced, these effects on opposite sides of the CME could cause deflection.  We focus only on the magnetic forces close to the Sun, ignoring magnetic reconnection.  

The model, called ForeCAT (Forecasting a CME's Altered Trajectory), calculates the deflection of a CME within a plane defined by global magnetic pressure gradients.  The deflection motion of the CME not only depends on the magnetic forces but requires models for the CME expansion and propagation as well.  ForeCAT uses the expansion model from the melon-seed-overpressure-expansion model (MSOE) of \citet{Sis06}.  A three-part propagation model, similar to \citet{Zha06}, determines the CME's radial motion.  The CME starts with a slow rise phase which transitions to an acceleration phase, then finally to a constant speed propagation phase.  ForeCAT also includes the effects of drag hindering the CME's nonradial motion, so that the CME cannot propagate freely in a direction quasi-perpendicular to the solar wind flow.  ForeCAT's radial propagation model results from fitting observations of CMEs affected by drag so ForeCAT does not explicitly include drag in the radial direction.

The paper is organized as follows: Section 2 contains analytic descriptions of the magnetic forces driving the deflection. Section 3 describes the expansion, propagation and drag models; section 4 presents two models for the background solar magnetic field: a scaled model and a Potential Field Source Surface (PFSS) model. Section 5 shows the results of a test case using the scaled magnetic background; section 6 investigates ForeCAT's sensitivity to input parameters, both CME parameters and values assumed in the analytic propagation model.  Section 7 includes results from using the PFSS magnetic field model.  Section 8 looks at deflection from local gradients related to the AR.

\section{Analytic Model of CME Deflection}
\subsection{Deflection Plane}
In order to simplify the treatment of the CME deflection in the lower corona, we restrict the calculations within ForeCAT to a plane called the ``deflection plane''.  In ForeCAT, magnetic forces drive the deflection so the background coronal magnetic field gradients at the location from which the CME launches determine the direction of the deflection plane.  The normal to this plane is defined as the cross product of the direction of initial radial CME motion and the direction of the dominant background magnetic pressure gradients. 

The calculation of the deflection plane normal vector uses the direction of the gradients in the magnetic pressure at a single location, which requires picking a specific height.  It is expected that the direction of the magnetic pressure gradient will change with distance from the Sun.  At smaller distances the local effects of the AR from which the CME is launched dominate the gradient, and at further distances effects from global features such as CHs and SBs dominate.  For the magnetic background used in this study, the effects from global features dominate at distances of 2 $\rsun$ or larger.  ForeCAT uses the direction of the gradient at this 2 $\rsun$ to define the deflection plane to capture the effects of the CH and SB.

Figure 1 shows four panels illustrating how different features determine the gradients at different heights.  The figures show a constant height from a MHD simulation using the Space Weather Modeling Framework (\citet{Tot11}, \citet{van10}, \citet{Eva12}, see section 4 for details) centered around the AR from which the CME is launched.  The figure contains color contours corresponding to the logarithm of the magnetic pressure and the arrows show the direction of the nonradial magnetic pressure gradient unit vectors.  The white dot indicates the latitude and longitude from which the CME launches.  All panels use the same color contour scale.  The strongest magnetic pressure occurs close to the AR, visible in Figure 1a (a distance of 1.05 $\rsun$).  Figure 1b shows a distance of 1.5 $\rsun$ where both the local effects of the AR and the global effects of the CH and SB influence the gradients.  In Figure 1c and 1d (distances of 2 $\rsun$ and 3 $\rsun$)  the streamer region that becomes the HCS can be seen as a minimum in the magnetic pressure.   In these panels the magnetic pressure is weaker than magnetic pressure in Figure 1a by 2-3 orders of magnitude.   At larger distances, the gradients transition from being dominated by local features, such as ARs, to a more uniform configuration, determined by the global structure of CHs and SBs.  For this background, 2 $\rsun$ is the smallest radius at which the gradients are dominated by the global effects.  These global gradients are present closer to the Sun, but can only be easily separated from the local gradients at larger distances.  Between 2 $\rsun$ and 3 $\rsun$ the direction of the gradient at the CME launch position changes by less than six degrees.  ForeCAT uses the value at the smaller radius where the magnetic field is stronger as more deflection will occur near that height.

\begin{figure}
\epsscale{1.0}
\plotone{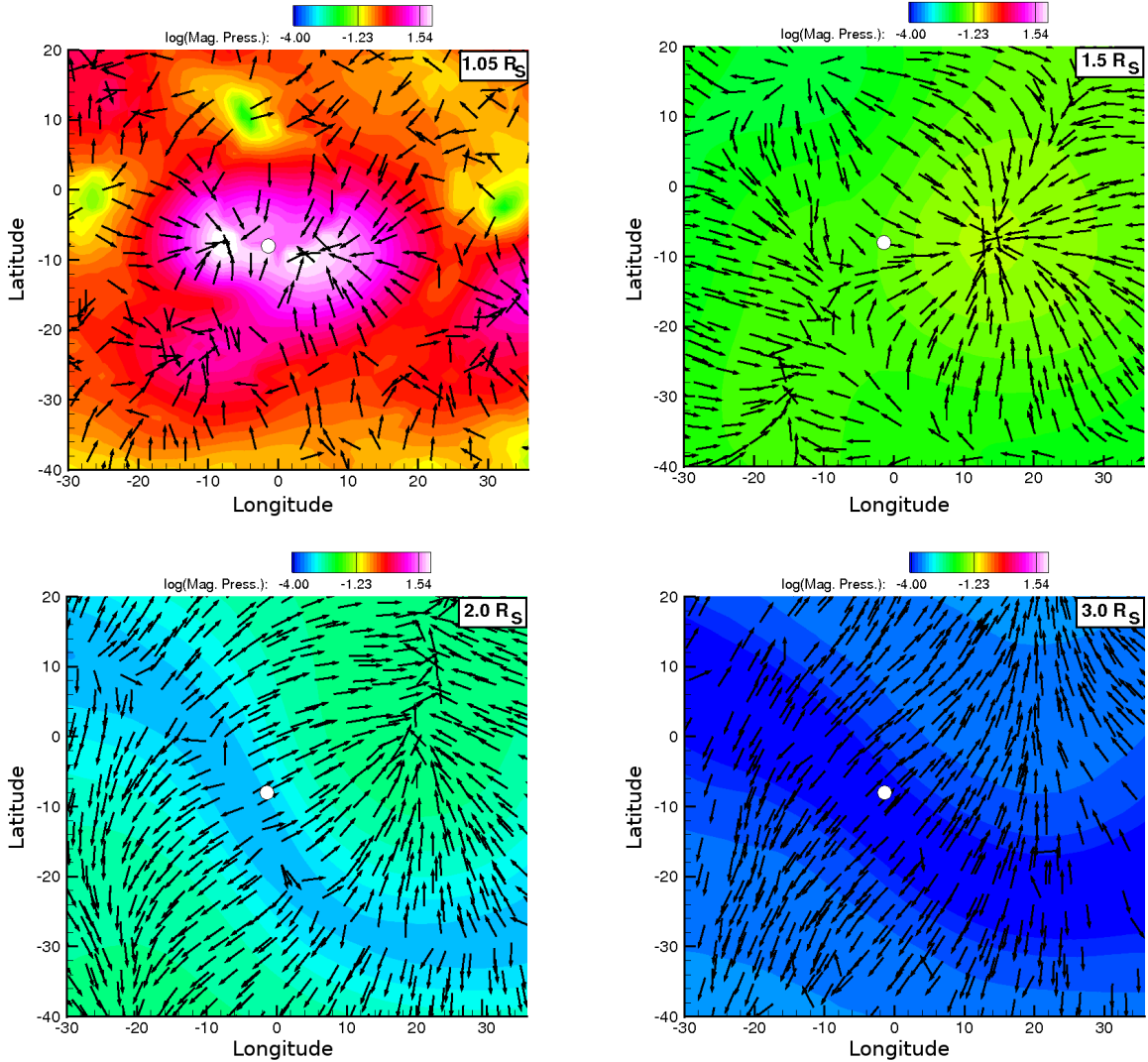}
\caption{All panels show latitude versus longitude at different heights, for a section around an active region.  The panels corresponds to distances a) 1.05 $\rsun$, b) 1.5 $\rsun$, c) 2.0 $\rsun$, and d) 3.0 $\rsun$.  The color contours show the magnetic pressure gradient, and arrows are unit vectors showing the direction of the magnetic pressure gradient in the plane.  The white dot indicates the launch position of the CME used in this study.  At low heights (R $<$ 2 $\rsun$) the AR dominates both the contours and gradients but as the distance increases, effects from the coronal hole and streamer belt become important.  In panels c and d the position of the streamer belt can be seen as a minimum in magnetic pressure.  Panel b shows an intermediate distance where both global and local effects influence the gradients.  A movie showing additional distances is available in the online version.}
\end{figure}

While the AR's magnetic field does affect the CME's propagation, and it is included in the calculations, the current focus of ForeCAT is deflection due to global gradients resulting from the orientation of CHs and SBs and the differences in these magnetic fields.  The effects of the AR are explored in section 8.  

The normal for the deflection plane is given by
\begin{equation}
\vec{n}=\vec{R}_0  \times -\vec{\nabla}\left(\frac{B^2}{8\pi}\right) 
\end{equation}
where $\vec{R}_0$ is the initial radial vector for the CME and $\vec{\nabla}B^2/8\pi$ is the gradient in the magnetic field pressure at a distance of 2 $\rsun$. 

Figure 2 illustrates how the deflection plane is selected.  Figure 2a shows color contours of the magnetic field strength at distances of 1.05 and 2 $\rsun$, analogous to Figure 1a and 1c, in three dimensions using a color scale appropriate for the range at each distance.  At 1.05 $\rsun$ the red lines indicate the approximate position of the CHs.  The black circle marks the latitude and longitude of the CME's initial position.  The radial vector $\vec{R_0}$ extends from the center of the Sun through this point. The black line shows the orientation of the deflection plane, defined by the direction of $-\vec{\nabla}B^2 / 8\pi$ at the black circle at 2 $\rsun$.  The radial direction and the gradient vector from Figure 2a define the deflection plane in Figure 2b.  The schematic in Figure 2b includes an example deflection plane and the Sun's surface.  As shown in Figure 2b, the deflection plane can be tilted, it need not be an equatorial or meridional plane. 

\begin{figure}
\epsscale{1.0}
\plotone{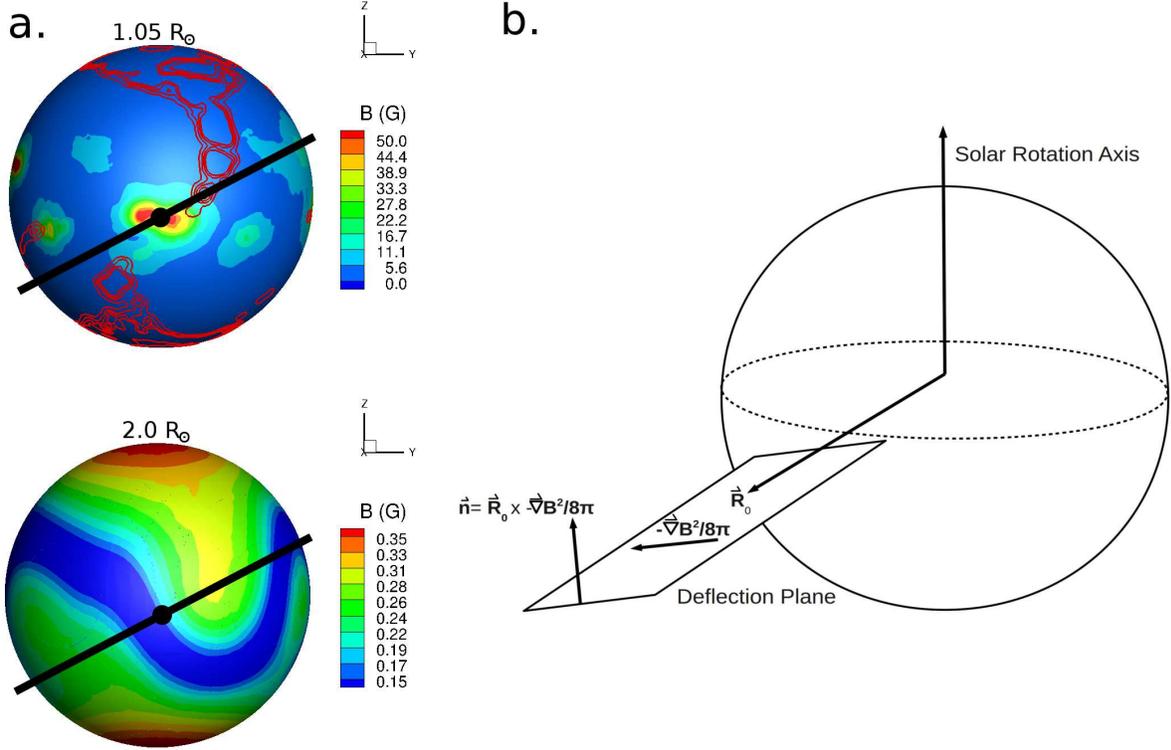}
\caption{Selection of the deflection plane.  Panel 2a shows the magnetic field strength at distances of 1.05 and 2 $\rsun$, similar to Figure 1a and 1c.  At 1.05 $\rsun$ red lines show the location of the nearby coronal holes.  The black dot indicates the initial latitude and longitude of the CME and the line shows the deflection plane orientation, the same as the magnetic pressure gradient direction at 2 $\rsun$ at the initial latitude and longitude.  The radial vector, $\vec{R_0}$ connects the center of the Sun to the black dot.  The normal of the deflection plane, $\vec{n}$ is defined in Equation 1.  Panel 2b shows the resulting deflection plane.}
\end{figure}

\begin{figure}
\epsscale{0.80}
\plotone{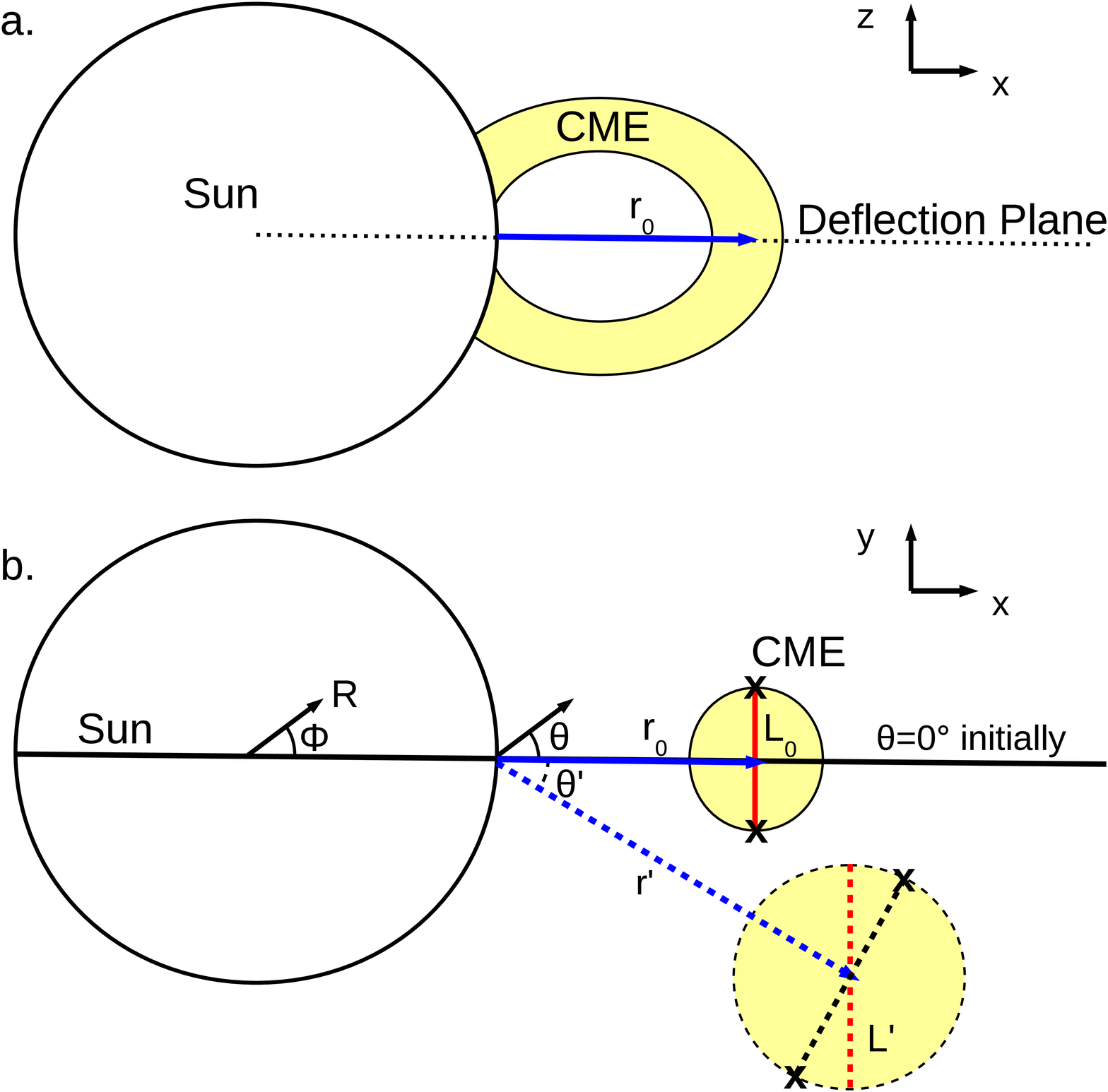}
\caption{Schematic showing details within the deflection plane.  Figure 3a shows the Sun, the flux-rope-like CME, and its intersection with the deflection plane.  In 3b the white circle represents the Sun and the yellow circle the cross-section of the CME flux rope within the deflection plane.  Two polar coordinate systems are shown as well as one Cartesian.  The x and y-directions correspond respectively to the $R_0$ and -$\nabla B^2/8\pi$ directions from Figure 2.  One set of polar coordinates ($r$ and $\theta$), used for the deflection force, is centered at the location from which the CME launches.  The Sun-centered polar coordinates ($R$ and $\phi$) are used to define a background magnetic field.  The circular CME cross-section starts with radius $L_0$, and position ($r$,$\theta$)=($r_0$,0).  The black X's mark the position of the CME edges where the deflection forces are calculated.  The red lines represent the diameter of the CME parallel to the y-axis which show the size and position of the CME in later figures.}
\end{figure}

Figure 3 contains a schematic showing the Sun-CME configuration: a) shows the CME, b) shows within the deflection plane.  Figure 3a shows the Sun in white and the flux-rope-like CME in yellow, as well as the intersection of the CME and deflection plane.  Figure 3b represents features within the deflection plane.  The background solar magnetic field is defined in polar coordinates, $R$ and $\phi$, with the origin at the center of the Sun.  A second set of polar coordinates, $r$ and $\theta$, with origin at the point on Sun ($ =1\rsun$) from which the CME is launched, is used to calculate deflection forces on the CME.  The set of Cartesian coordinates, with origin also at the center of the Sun, allow conversion between the two sets of polar coordinates:
\begin{displaymath}
 R=\sqrt{x^2+y^2} \qquad r=\sqrt{(x-1)^2+y^2}
\end{displaymath}
\begin{displaymath}
\phi=\tan^{-1}\left(\frac{y}{x}\right) \qquad \theta=\tan^{-1}\left(\frac{y}{x-1}\right)
\end{displaymath}
where $x$, $y$, $R$, and $r$ have units of $\rsun$ and $x$ and $y$ correspond respectively to the $\vec{R_0}$ and $\vec{\nabla}B^2/8\pi$ directions in Figure 2.  In this geometry the CME launches along the x-axis which corresponds to $\theta=\phi=0$. 

A circle, initially of radius $L_0$, represents the cross-section of the CME in the deflection plane.  For the cross-section to be a perfect circle requires the CME to be perpendicular to the deflection plane.  Deviations from this orientation will introduce small errors into the calculation of the CME density as the cross-section will take an elliptical shape within the deflection plane.  ForeCAT uses the deflection forces on two ``edges'', marked with X's in Figure 3, to calculate the total deflection.  These edges correspond to the points on the circle that lie on a line running through the center of the circle and perpendicular to the $\hat{r}$ direction at that point.  Averaging the $\phi$ values of the two edges gives the central position angle (CPA) of the CME, which equals the $\phi$ position of the center of the circle: 
\begin{equation}
CPA=\frac{\phi_{1}+\phi_2}{2}
\end{equation}
where 1 and 2 refer to the two edges. The CPA is calculated using the change in the Sun-centered angle, comparable to latitudinal or longitudinal changes of CMEs in coronagraph observations.

The net deflection force on the two edges determines the change in the $\theta$ position of the CME ($\theta\rightarrow\theta'$).  Before a CME detaches from the solar surface the deflection motion will occur with respect to the position where the footpoints are anchored. Accordingly, ForeCAT calculates deflection forces in the $\hat{\theta}$ direction.  Different analytic models describe the change in distance ($R\rightarrow R$'), change in CME radius ($L \rightarrow L$'), and effects of nonradial drag, separate from the deflection (see section 3).  No change occurs in the CPA for a CME propagating without deflection and with uniform expansion.  Deviations from the original CPA correspond to deflection or non-uniform expansion, however ForeCAT only includes CMEs with uniform expansion (section 3.2).

\subsection{Deflection Forces}
ForeCAT calculates CME deflection due to the magnetic tension and the magnetic pressure gradient.  Imbalance of these forces between the two edges causes a net force in the $\hat{\theta}$ direction, driving deflection.  All forces within this model are volumetric so that the acceleration equals the force divided by the density.

\subsubsection{Magnetic Tension}
In general the force due to magnetic tension can be expressed as
\begin{equation}
F_{\kappa}=\vec{\kappa}\frac{B^2}{4\pi}
\end{equation}
where $\kappa=1/R_C$ is the curvature and $R_C$ is the radius of curvature.  The tension force points toward the center of curvature.  As the CME expands into the surrounding medium the external magnetic field will drape around it.  The curvature of the draped magnetic field can be approximated then as equal to the CME curvature with $R_C$ as the radius of the CME cross-section within the deflection plane.

The draping of the coronal magnetic field is not restricted to the deflection plane so ForeCAT includes a $\cos \alpha$ factor to account for this, assuming the radius of curvature does not change.  The angle $\alpha$ is the angle between the deflection plane and the direction of the draping of the background solar magnetic field lines around the CME.  In principle $\alpha$ will vary in time.  The final tension force on each edge is
\begin{equation}
\vec{F}_{\kappa}=\mp\frac{1}{L}\frac{B^2}{4\pi}\cos\alpha\;\hat{\theta}
\end{equation}
where the top edge (defined as the edge with the largest $y$ value in the Cartesian coordinate system in Figure 3b) has the negative sign and the bottom the positive.  Only for a background magnetic field that is symmetric about the CME will the two forces balance.

\subsubsection{Magnetic Pressure Gradient}
The component of the magnetic pressure gradient perpendicular to the radial direction also leads to deflection: 
\begin{equation}
F_{\nabla P}=-\nabla_{\perp}\frac{B^2}{8\pi}
\end{equation}
where the $\perp$ corresponds to gradients perpendicular to the direction of the magnetic field according to the definition of the Lorentz force.  The magnetic pressure gradient expression used in ForeCAT includes the $\cos\alpha$ factor to account for draping out of the deflection plane.  Since the background magnetic field lines drape around the CME, at the edges the direction of the perpendicular gradient within the deflection plane is the $\hat{\theta}$ direction.  Equation 5 can be recast as 
\begin{equation}
\vec{F}_{\nabla P}=-\frac{B}{4\pi R}\frac{\partial B}{\partial \phi}\cos(\theta-\phi)\cos\alpha\;\hat{\theta}
\end{equation}
with the force directed in the $\hat{\theta}$ direction due to the $\cos(\theta-\phi)$ term which results from taking the $\hat{\theta}$ component of the gradient in $\phi$.  As the CME propagates away from the Sun the orientation of the background magnetic pressure gradients may change.  When this occurs deflection occurs out of the original deflection plane.  The net out of plane deflection is minimal as the magnetic forces decrease with distance.   

\subsubsection{Total Deflection Force}
The net volumetric deflection force is given by the sum of equations 4 and 6.
\begin{equation}
\vec{F} = \left(\mp\frac{1}{L}\frac{B^2}{4\pi}\cos\alpha-\frac{B}{4\pi R}\frac{\partial B}{\partial \theta}\cos(\theta-\phi) \cos\alpha\right) \hat{\theta}
\end{equation}
where $\hat{\theta}=-\sin\theta\hat{x}+\cos\theta\hat{y}$, which changes with time as $\theta$ changes.  Dividing the deflection force by the CME density gives the acceleration of each CME edge.  The density is defined as: 
\begin{equation}
\rho=\frac{M}{\pi^2 r L^2}
\end{equation}
which $M_{CME}$ is the CME mass and the volume is approximated using a uniform curved cylinder of length $\pi r$ and cross-section $\pi L^2$. The mass of the CME is assumed to be constant.  CMEs tend to accrete mass as they travel.  \citet{Vou10} analyze the mass evolution of CMEs in the low corona using the coronagraph brightness and find that CMEs tend to increase in mass in the corona below 10 $\rsun$.  ForeCAT's assumption of a constant mass will cause underestimations of the density, leading to an overestimate of the acceleration cause by deflection. The density evolves in time due to the expansion of the CME (see section 3.1). 

The acceleration of edges in the $\theta$ direction is a linear acceleration with $x$ and $y$ components.  The equations of motion for an edge are:
\begin{eqnarray}
x(t+\Delta t) &=& x(t)+v_{def,x}(t)\Delta t-0.5\frac{F(t)}{\rho(t)}\sin\theta (t) \Delta t^2\\ 
y(t+\Delta t) &=& y(t)+v_{def,y}(t)\Delta t+0.5\frac{F(t)}{\rho(t)}\cos\theta (t) \Delta t^2 \nonumber
\end{eqnarray}
where $v_{def,x}$ and $v_{def,y}$ are the velocities of the edge in the $\hat{x}$ and $\hat{y}$ direction resulting from deflection
\begin{eqnarray}
v_{def,x}(t)&=&-\int^t_0 \frac{F(t)}{\rho (t)} \sin\theta (t) dt\\
v_{def,y}(t)&=&\quad\int^t_0 \frac{F(t)}{\rho (t)} \cos\theta (t) dt \nonumber
\end{eqnarray}  
The deflection equals the change in the CME's CPA  
\begin{equation}
CPA(t)=\frac{1}{2}\left[\tan^{-1}\left(\frac{y_1(t)}{x_1(t)}\right)+\tan^{-1}\left(\frac{y_2(t)}{x_2(t)}\right)\right]
\end{equation}
where the 1 and 2 refer to the two CME edges.  Since initially the CPA equals zero, the total deflection at any time equals the CME's current CPA.  The total deflection within the deflection plane can be converted into a change in latitude and longitude using the orientation of the plane.

\section{Description of CME Motion}
To calculate the total deflection, the radial propagation, expansion, and nonradial drag of the CME must be incorporated as they affect the position of the CME edges over time.  

\subsection{CME Expansion}
Several analytic models describing CME evolution exist.  These models focus on the radial propagation of CMEs and do not account for deflections.  \citet{Pne84} introduces the melon-seed model, which \citet{Sis06} later develop into the melon-seed-overpressure-expansion (MSOE) model.  More complex models exist such as that of \citet{Che96}, which treats the CME as a flux rope containing two different plasmas, representing the cavity and embedded prominence, and triggers the eruption by increasing the poloidal magnetic field.  ForeCAT uses the MSOE model's description of CME expansion.

The MSOE model modifies a classical hydrodynamic solution for an overpressure of a spherical cavity.  The hydrodynamic solution (see \citet{Mil68}) is driven by an adiabatic gas overpressure that can be treated as a fluid ``source.''  \citet{Sis06} change the adiabatic overpressure to a magnetic overpressure.  This results in the following expression for the change in the CME radius L (see Appendix for more details).
\begin{equation}
\frac{\partial^2L}{\partial t^2}=\frac{1}{L}\left[-\frac{3}{2}\left(\frac{dL}{dt}\right)^2 + \left(\frac{L_0}{L}\right)^4\frac{(A_{h0}^2-A_{SW0}^2)}{2}\frac{\rho_{SW0}}{\rho_{SW}}\right]
\end{equation}
The subscript $0$ indicates initial values evaluated at t=0.  $A_{h0}$ and $A_{SW0}$ refer to Alfv\'en speeds calculated using the initial background solar wind density and either the magnetic field of the CME ($A_{h0}$) or the background solar magnetic field strength ($A_{SW0}$).  The expansion equation depends on the background solar wind density $\rho_{SW}$ which requires assuming some solar wind density profile.  

ForeCAT uses the expression for the density from \citet{Che96}, also used by \citet{Sis06}:
\begin{equation}
\rho_{SW}(R)=6.68\times10^{-16}\left[3\left(\frac{\rsun}{R}\right)^{12}+\left(\frac{\rsun}{R}\right)^4\right]+3.84\times10^{-19}\left(\frac{\rsun}{R}\right)^{2} \mathrm{g} \; \mathrm{cm}^{-3}
\end{equation}
Using the profile from \citet{Che96}, the CH regions are scaled down by a constant value as $\rho_{CH}(R)=0.25\rho_{SW}(R)$, which produces a CH density profile closer to that of observations \citep{Guh98,Doy99}.  The value 0.25 results from assuming constant mass flux and a solar wind speed roughly double the slow wind for the fast wind \citep{Mcc00}.  

Figure 4 shows the analytic density model and several radial profiles from a MHD simulation, the same simulation shown in Fig. 1 and 2.  The details of the simulation are discussed in section 4.1.  The MHD profiles come from different locations: above an AR (red), a CH (blue), and the SB (green).  The standard analytic model (solid black) is shown in addition to the scaled CH analytic model (dashed black).  Close to the Sun the analytic model overestimates the MHD solution by nearly an order of magnitude.   Near 20 $\rsun$, the outer boundary of the MHD simulation domain, the analytic and MHD profiles for the SB and CH have better agreement.  However, some discrepancy still exists for the AR.  The effects of the chosen analytic density profile will be explore in a future work. 

\begin{figure}
\epsscale{1.0}
\plotone{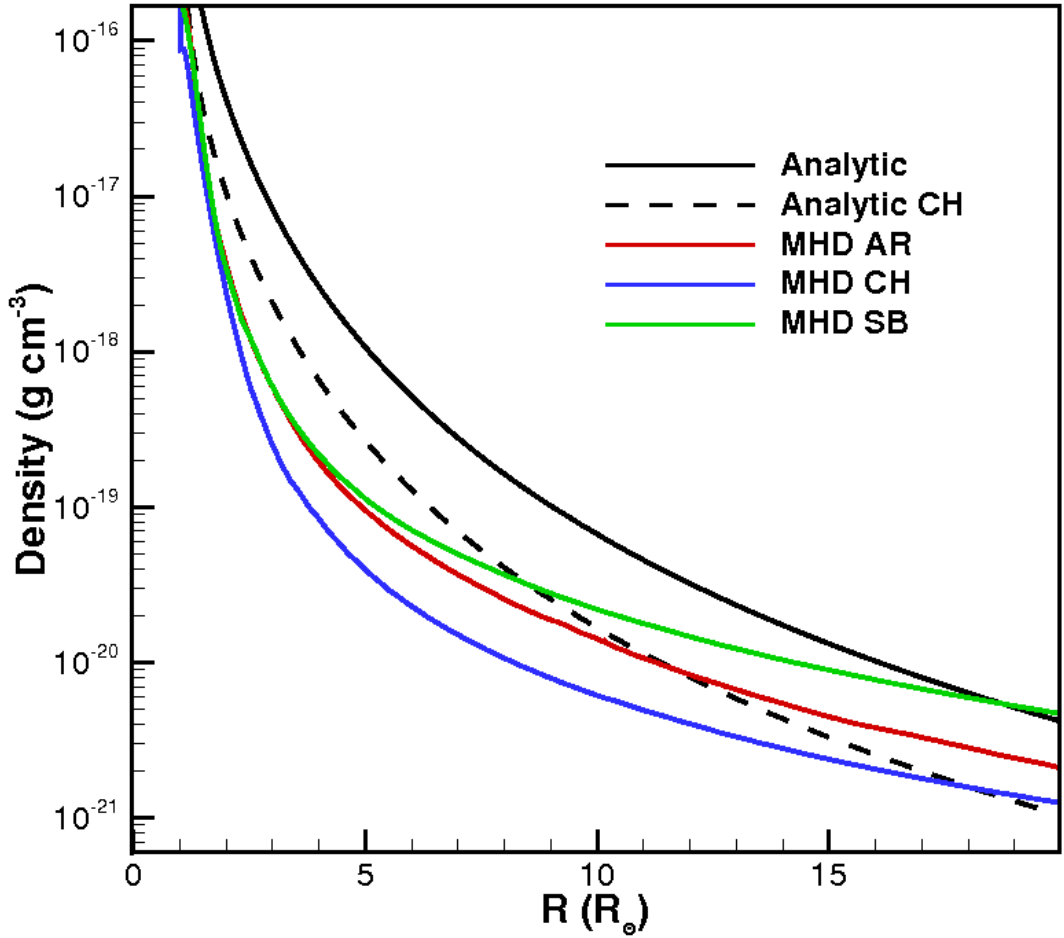}
\caption{Radial density profiles for the analytic model and the results of the MHD simulation.  The analytic model overestimates the MHD profiles from above an active region (red), coronal hole (blue), and the streamer belt (green).  Near the edge of the MHD domain better agreement exists between the models.}
\end{figure}

\subsection{CME Propagation}
ForeCAT adopts an analytic expression for the radial propagation speed based on empirical fits.  The Lorentz force that drives the CME deflection also causes radial motion of the CME, however, ForeCAT makes use of an empirical model.  \citet{Zha01,Zha04} and \citet{Zha06} present a three phase description of CME propagation (initiation or gradual rise, impulsive acceleration, and propagation) and the connection to X-ray flare observations.  The initiation phase occurs first as the CME slowly rises, then due to an instability or reconnection the CME lifts off and begins rapidly accelerating away from the Sun.  The transition to the acceleration phase often correlates with the onset of flare activity.  Both the initiation and propagation phases occur within a few solar radii of the Sun \citep{Zha06}.  The final phase is the propagation phase where comparatively little CME acceleration occurs.  In a statistical study of 50 CMEs, \citet{Zha06} found average main accelerations of 330.9 m s$^{-2}$ whereas the average residual acceleration during the propagation phase was only 0.9 m s$^{-2}$.  

\citet{Vrs07} and \citet{Bei11} identify the Lorentz force as the driving mechanism behind the impulsive acceleration in the radial direction, explaining why this phase occurs so close to the Sun's surface.  The observational studies of both \citet{Bei11} and \citet{Jos11} show that the acceleration phase tends to end by a height of about 2 $\rsun$.

Although the acceleration phase contains the most rapid acceleration, a CME continues to accelerate during the propagation phase, \citet{Che10} refer to this as the post-impulsive-phase acceleration.  The \citet{Che10} study of several hundred CMEs results in a mean post-impulsive-phase acceleration value equal to -11.9 m s$^{-1}$ with individual values ranging between -150 and 180  m s$^{-1}$.  The data from \citet{Zha06} cover a similar range of post-impulsive-phase accelerations as the data from \citet{Che10} but the two means differ as a result of using different subsets of CMEs from the LASCO catalog.  The positive and negative post-impulsive-phase acceleration values imply that CMEs can either accelerate or decelerate in the third phase due to the drag force from the CME's interaction with the solar wind.  

ForeCAT adopts the three part propagation model for the radial dynamics.  ForeCAT uses a constant velocity for the initiation and propagation phase and a constant acceleration in the acceleration phase.  We define the radial distance at which the CME transitions from the gradual to acceleration phase as $R_{ga}$ and that from the acceleration to propagation phase as $R_{ap}$.  We assume that the initiation phase lasts until the center of the CME cross-section reaches a distance $R_{ga}=1.5\rsun$  and then the acceleration occurs until $R_{ap}=3.0\rsun$.  ForeCAT uses a single representative value for each transition, as well as a constant value for the gradual velocity of the CME in the initiation phase, $v_g$=80 km s$^{-1}$.  \citet{Zha06} observe $v_g$ between tens of km s$^{-1}$ up to 100 km s$^{-1}$.  Section 6 contains analysis of ForeCAT's sensitivity to the parameters $R_{ga}$, $R_{ap}$, and $v_g$. 

Given the above assumptions, the CME's radial propagation is described by its final velocity, $v_f$ at the propagation phase.  From the kinematic evolution of the CME during acceleration phase
\begin{equation}
v_f^2=v_g^2+2a(1.5\rsun)
\end{equation}
which corresponds to an acceleration, a,
\begin{equation}
a=\frac{v_f^2-v_g^2}{3 \rsun}
\end{equation}
 which allows us to describe the CME's radial velocity over time as
\begin{eqnarray}
v_{r}&=&v_g \qquad 1.0 \rsun\leq R \leq R_{ga} \\ \nonumber
v_{r}&=&v_g+0.5a(t-t_{ga}) \qquad R_{ga} \le R \leq R_{ap} \\ \nonumber
v_r&=&v_f \qquad  R \ge R_{ap} 
\end{eqnarray}
where t$_{ga}$ is the time at which the CME reaches R$_{ga}$.  Equation 16 produces CME velocity profiles similar to those present in Figure 1 of \citet{Zha06}, with the exception of flat initial and propagation phases.

\subsection{Nonradial Drag}
We include the nonradial drag as the component of the drag force in the $\hat{\phi}$ direction which results from the interaction of CME with the solar wind.  ForeCAT does not explicitly calculate drag in the radial direction since the propagation model describes a CME's radial motion.  To calculate the nonradial drag, ForeCAT uses the expression for the acceleration due to drag from \citet{Car04}:
\begin{equation}
\vec{a}_D=-\frac{C_dA\rho_{SW}}{M_{CME}}(\vec{v}_{CME}-\vec{v}_{SW})|\vec{v}_{CME}-\vec{v}_{SW}|
\end{equation}
where $C_d$ is the drag-coefficient, $A$ the cross-sectional area in the direction of the drag, and $v_{SW}$ the solar wind speed.  \citet{Car04} use this equation to describe the radial drag on a CME, but the same physical process governs drag in all directions.  The solar wind is approximated as entirely radial so that the solar wind velocity term equal zero in the expression for the drag in the $\hat{\phi}$ direction.  Close to the Sun this approximation is the least accurate but it allows ForeCAT to include nonradial drag without invoking a complete solar wind velocity model.

The cross-sectional area in the direction of the drag can be expressed as:
\begin{equation}
A=\frac{\pi}{2}\left((r+L)^2-(r-L)^2\right)=2\pi rL
\end{equation}
based on the definition of the CME structure in section 2.  Setting $C_d$=1 and taking the component of the CME velocity in the radial direction, equation 17 becomes:
\begin{equation}
\vec{a}_d=-\frac{2\pi rL\rho_{SW}}{M_{CME}}\left(-v_x\sin(\phi)+v_y\cos(\phi))\right|-v_x\sin(\phi)+v_y\cos(\phi)|\hat{\phi}
\end{equation} 
where $v_x$ and $v_y$ are the x and y components of the CME's velocity.  In section 6 we explore ForeCAT's sensitivity to the drag coefficient and discuss models for $C_d$.

\section{Background Solar Magnetic Field}
To calculate a CME's deflection, ForeCAT includes the magnetic structure of various features such as CHs, ARs, and SBs.  The background solar magnetic field therefore is crucial.  Two different background magnetic field models were explored with ForeCAT: a ``scaled'' background and a Potential Field Source Surface (PFSS) background.

\subsection{Scaled Background}
The scaled background uses the background magnetic field from the output of an MHD steady state solar wind from the Space Weather Modeling Framework (SWMF, \citet{Tot12}, \citet{van10}) using a magnetogram as input.  Alfv\'en waves drive the background solar wind and surface Alfv\'en wave damping adds heating \citep{Eva12}.  The magnetic field values from a ring at $R=1.15\rsun$ within the deflection plane (defined using the magnetic pressure gradients at a distance of 2 $\rsun$) within $\pm 90\mydeg$ of the CME launch location yield discrete points for the magnetic field strength as a function of angle.  By extracting values at low heights, $B(\phi )$ includes the signatures of the solar features (CHs, SBs and ARs).  ForeCAT uses the MHD background only at $R=1.15\rsun$ and uses extrapolations for larger radii based on observational studies of the solar magnetic field.  ForeCAT uses these extrapolations because, as described below, the MHD solution falls unrealistically quickly.  The extrapolations differ between ARs and non-AR locations, also described below.

The extrapolations for ForeCAT's magnetic field model result from observations of the solar magnetic field versus distance.  Observational studies of the magnetic field of ARs fit the profile of the magnetic field versus distance with the form $B=B_0R^{-\alpha}$.  \citet{Dul78} present a compilation of observational data(including data from Helios, Mariner 10, and various ground-based solar telescopes) of the magnetic field above an active region.  The study finds that $B=0.5[(R/\rsun )-1]^{-1.5}$ agrees within a factor of three for all the observations.  \citet{Pat87} use Helios measurements of Faraday rotation and find a best fit between 3 and 10 $\rsun$ using a combination of $\alpha=2$ and $\alpha=3$.  More recent Faraday rotation measurements have been acquired for $R$ between 6.2 and 7.1 $\rsun$ using the Very Large Array which agree with a coefficient of $\alpha$=1.3 \citep{Spa05} .  In order to study shock development in the corona, \citet{Man03} use a background magnetic field combining a $R^{-2}$ term for the quiet sun and a dipole term ($\propto \; R^{-3}$) to represent the ARs. \citet{Gop11} use the standoff distance of CME-driven shocks to determine the magnetic field profile between 6 and 23 $\rsun$ and find good agreement with \citet{Dul78} and \citet{Spa05}.

Fitting polynomials to these points allows generalization of the discrete magnetic field points to a function that can be used for all $\phi$ angles. In addition, it allows for calculation of analytic derivatives.  Separate polynomials describe the AR and the quiet sun (QS, defined as the region outside of the AR).  The ranges of the polynomials are determined by the location of local maxima and minima in $B(\phi )$, having the polynomials break at inflection points provides the best fit.   First polynomials are fit to the QS yielding a function $B_{QS}(R,\phi)$.  The QS magnetic field is subtracted from the MHD result and then the AR polynomials, $B_{AR}(R,\phi)$ , are fit to the residual magnetic field.  Figure 5a shows the simulation data (solid red line) as well as the sum of the QS and AR best-fit polynomials (dashed black line) for $B(R=1.15\rsun,\phi)$ for $-90\mydeg\le\phi\le 90\mydeg$. 

\begin{figure}
\epsscale{1.0}
\plotone{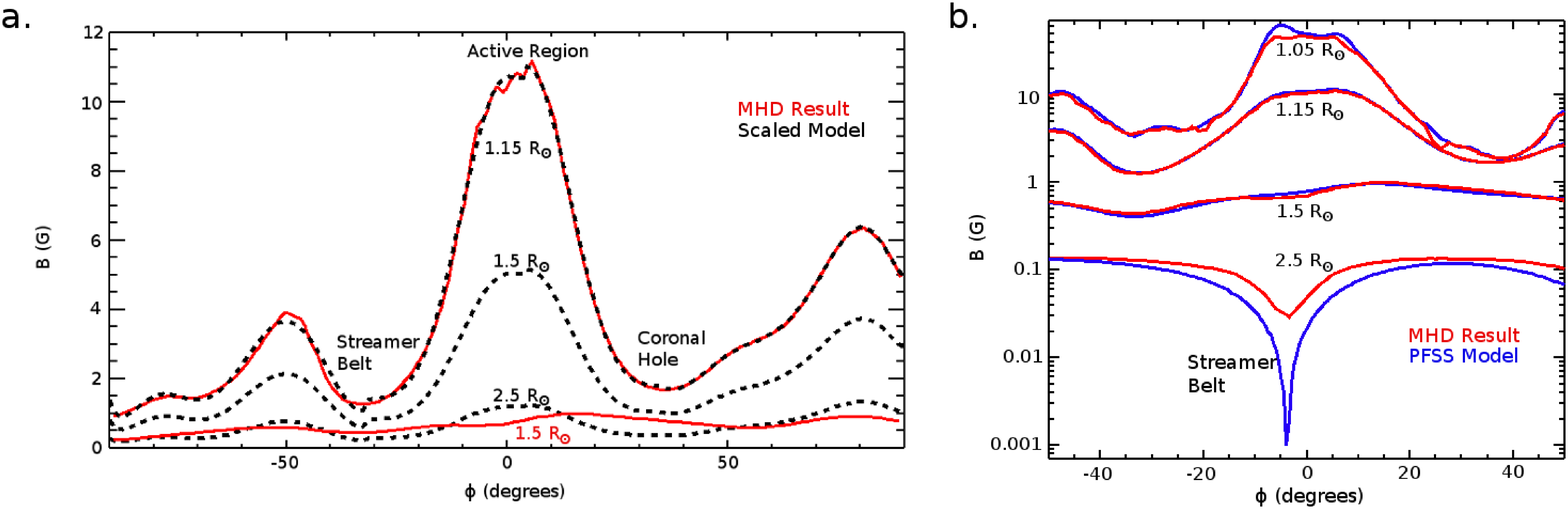}
\caption{The left panel contains the scaled model background magnetic field at different heights in the deflection plane.  The dashed black lines represent the scaled model for the angular dependence of the magnetic field strength between 1.15 $\rsun$ and 2.5 $\rsun$ and the solid red lines the results from the MHD simulation.  There are labels for solar features such as a coronal hole (CH), an active region and the streamer belt (SB) which correspond to their location at 1.15 $\rsun$.  Asymmetry between the magnitude of $B$ at the minima corresponding to the SB and CH causes a global gradient that drives the CME deflection.  The MHD model magnetic strength falls quicker with distance than the scaled model.  The right panel shows the PFSS model in blue and the MHD in red.  In panel b, both models show the angular magnetic profile changing with distance; the minima at 1.15 $\rsun$ are not defined to be the minima at further distances.}
\end{figure}

Within the deflection plane, the AR corresponds to two local maxima in $B(\phi)$ whereas both the SB and CH represent local minima. The center of the AR is a local minimum corresponding to the polarity inversion line (PIL) located in between the two maxima corresponding to the opposite polarity flux systems.  Weaker magnetic field exists at the SB minimum than the CH minimum and this asymmetry produces the gradients that drive deflection.  This example contains strong gradients due to the proximity of the CH and SB.  Any coronal configuration will have gradients leading to deflection, the magnitude of the deflection depending on the magnitude of the gradients. 

Scaling the values from $R=1.15\rsun$ determines the background magnetic field strength at other radii.  ForeCAT's model treats the AR like a dipole so that the magnetic field falls as $R^{-3}$.  Outside the AR, for the QS the model uses the $R^{-2}$ dependence commonly used for open field lines.  This combination of scaling is the same as that of \citet{Man03}. Equation 20 gives the ForeCAT scaled magnetic field.
\begin{equation}
B (R, \phi) = B_{QS}(1.15 \rsun, \phi) \left(\frac{1.15 \rsun}{R}\right)^2 +B_{AR}(1.15 \rsun, \phi) \left(\frac{1.15 \rsun}{R}\right)^3
\end{equation}

\citet{Man03} use type II radio bursts to infer the behavior of the background solar magnetic field.  The type II radio bursts are believed to result from shock waves propagating outward in the corona \citep{Nel85}.  The speed of the disturbance driving the shock can be used to infer the background Alfv\'en speed and because the type II emissions occur at the local background plasma frequency the Alfv\'en speed can be used to determine the background magnetic field strength.  \citet{Man03} compare the Alfv\'en profile from their magnetic field model with the general behavior of type II radio bursts.  The combination of a scaling of $R^{-2}$ and $R^{-3}$ yields favorable comparisons to the type II radio observations.  In particular, the model produces an Alfv\'en profile with a local minimum and maximum in the low corona which allows for the formation, decay, and reformation of shocks within 6 $\rsun$, reproducing a two shock wave behavior seen in some type II radio observations \citep{Gop02}.

Figure 5a shows the scaled model magnetic field strength for several different radii ($R$=1.5, 2.0, and 2.5 $\rsun$).  As the radius increases the signatures of the individual solar features weaken but are still present at 2.5 $\rsun$.  The MHD results for $R$=1.5 are also included.

ForeCAT uses analytic fits to observations rather than the results of MHD simulations because for $R<2\rsun$ , as seen in Figure 5a, the MHD magnetic field strength decreases quicker with distance than observations, closer to $r^{-6}$ or $r^{-8}$ depending on the region (CH or SB versus AR).  Recent advancements in the MHD model have included a two-temperature (electron and proton) formalism, including the effects of field-aligned heat conduction, radiative cooling, collisional coupling, and wave heating \citep{Dow10, van10, Sok13}.  In addition, a Finite Difference Iterative Potential Solver (FDIPS; \citet{Tot11}) can be used to initialize the magnetic field in place of the spherical harmonic expansion approach.  However, the wave-driven model output (with either FDIPS or harmonic coefficients) does not show a difference in the rapid decrease of the magnetic field magnitude with radial distance. 

Beyond 2 $\rsun$ the MHD magnetic field falls as the expected r$^{-2}$, but, as discussed in Section 5, the deflection of the CME depends crucially on the magnetic strength at these distances.  The rapid decrease of the magnetic field in MHD simulations will lead to an underestimate of the magnetic deflection,  which could explain the discrepancy between the observed and simulated CME in \citet{Lug11}.  \citet{Eva08} show that the steepness of the MHD profiles would allow slow CMEs to drive shocks low in the corona and that the Alfv\'en speed profiles do not have the characteristic ``valley'' and ``hump'' shape seen in analytic models.  Using a scaled model, we capture a slower decrease of B with distance, consistent with some type II radio observations.  However, this model does not allow a change in the angular position with distance of coronal structures such as the SB.

\subsection{PFSS Background}
PFSS models were first used to describe the solar magnetic field in the late 1960's \citep{Alt69,Sch69}.  If the magnetic field is assumed to be potential, it can be described using a sum of Legendre polynomials.  The harmonic coefficients can be determined from a magnetogram and by assuming the magnetic field becomes entirely radial at the source surface height.  The magnetic field at any location can be calculated using the harmonic coefficients.  The literature contains extensive discussion of the details of PFSS calculations and the model's ability to reproduce observed conditions \citep{Hoe82,Luh02,Neu98,Ril06,Wan92,Wan93}.

ForeCAT uses a PFSS magnetic field strength calculated using radial harmonic coefficients \citep{Wan92} from the Michelson Doppler Imager on SOHO \citep{Sch95}, calculated with a source surface height of 2.5 $\rsun$.  The PFSS magnetic field is calculated using coefficients for Legendre polynomials up to order 90.  Higher order polynomials represent spatially smaller features and decay faster with distance.  Since the magnetic field strength (which drives ForeCAT deflection) is strongest close to the Sun, not including the higher orders could make a difference in the CME deflection.  Figure 5b contains the PFSS magnetic field (in blue), as well as the MHD results (in red) for $R$ = 1.05, 1.15, 2.0 and 2.5 $\rsun$ within $\pm 50 \mydeg$ of the location from which the CME launches.

In Fig. 5b the PFSS model and the MHD model agree.  \citet{Ril06} find similar agreement between the MHD and PFSS solutions when strong currents are not present.  Both models also show a clear change in the angular magnetic field profile with distance, an effect that the scaled model cannot capture.  The ``rigid'' magnetic minima of the scaled model exists at 1.15 $\rsun$ but at 2.5 $\rsun$ the formation of the HCS near -5$\mydeg$ causes a different magnetic minima.  This change of the magnetic minima will affect the CME's magnetic deflection.  The PFSS model and the MHD model fall similarly with distance.  The PFSS background will underestimate magnetic forces compared to the scaled background.

\section{Numerical Implementation and Test Case}
Equations 9, 12 and 16 form a set that describes the evolution of the CME as it propagates away from the Sun for the model, ForeCAT.  Initializing the equations requires a radius of the cross-section of the CME within the deflection plane, $L_0$, height, $r_0$, the CME mass, $M_{CME}$, the final propagation velocity, $v_g$, and the magnetic field strength of the CME that causes the initial overpressure, $B_0$.  We assume that the angle $\alpha$ equals zero throughout the simulation (no draping outside of the deflection plane) and therefore find a maximum deflection angle.  The model also needs the background magnetic field configuration.

ForeCAT integrates these equations numerically using a second-order Taylor expansion for the position so that the error is of order $\Delta t^3$.  ForeCAT yields a deflection of the CPA over time, as well as the trajectory of the CME as it deflects. 

In the control case, the following values are chosen for the free parameters of ForeCAT: $M_{CME}$=10$^{15}$ g, $v_{g}$=475 km s$^{-1}$, $L_{0}$=0.15 $\rsun$ and $B_0$= 15 G.  These input parameters represent an average CME mass and the velocity corresponds to the mean value from the \citet{Gop09a} analysis of the SOHO/LASCO survey of CMEs before the end of 2006.  \citet{Sis06} use a similar value of CME magnetic strength.  The CME begins at a height of 0.25 $\rsun$ so the model captures some of the gradual phase of radial propagation.  The CME launches from AR 0758 of Carrington Rotation (CR) 2029.  This corresponds to the magnetic field background shown in Figure 5, this case using a scaled magnetic background.  The deflection plane was defined using the magnetic pressure gradient at $R$=2$\rsun$ and a latitude of -8$\mydeg$ and Carrington longitude of 130.6$\mydeg$.  Strong gradients that exist between the SB and CH should cause a large deflection.  Figure 6a shows the CME's propagation out to a distance of about 10 $\rsun$ in a Cartesian coordinate system with the Sun at the origin. The figure shows the diameter of the CME parallel to the y-axis (shown with a red line in Figure 3) in 1 minute time-steps.  Figure 6b shows the CPA (Eq. 11) of the CME versus distance out to 1 AU.

\begin{figure}
\epsscale{1.0}
\plotone{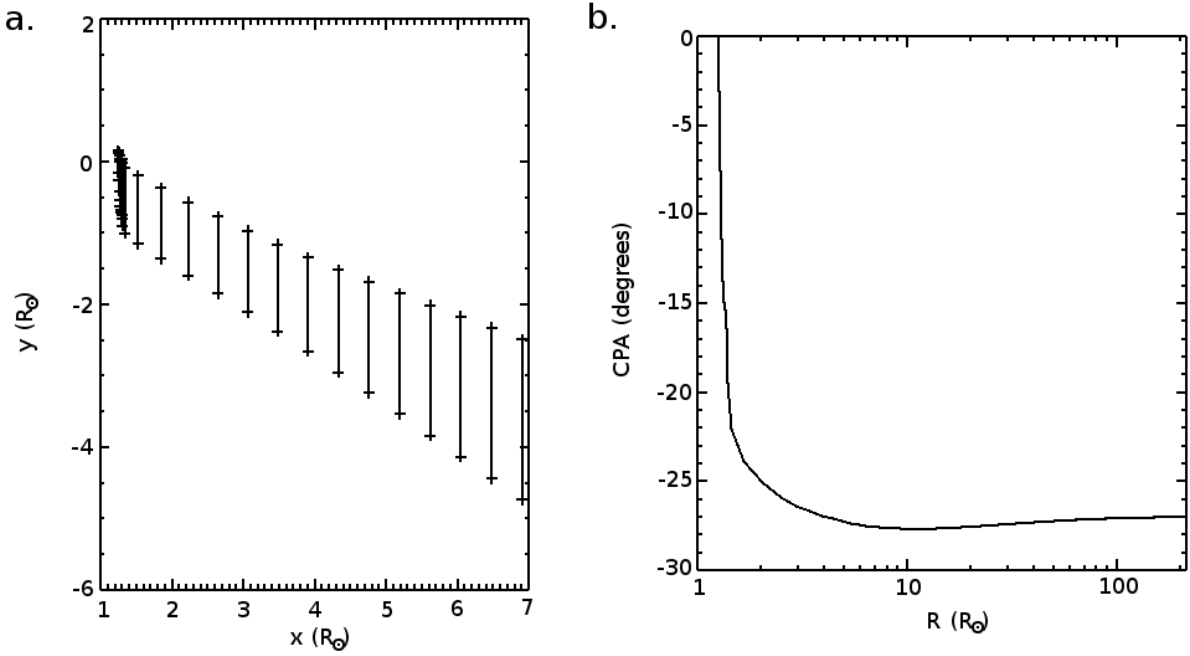}
\caption{The ForeCAT results for the deflection of the control case ($M_{CME}$=1x10$^{15}$ g, final propagation speed $v_f$=475 km s$^{-1}$, CME magnetic field strength $B_{0}$ = 10 G, initial radius $L_0$=0.15 $\rsun$ and initial distance $r_0$=0.25$\rsun$).  The model yields a total deflection of -27.0$\mydeg$ which corresponds to a change of -8.1$\mydeg$ in latitude and -26.4$\mydeg$ in longitude.  Panel a shows a subsection of the trajectory close to the Sun, highlighting the deflection within $R< 10 \rsun$ by showing the diameter of the CME cross-section parallel to the y-axis (the red line in Figure 3).  Panel b shows the evolution of the CPA of the CME out to 1 AU.}
\end{figure}

The CME deflects -27.0$\mydeg$ in the deflection plane during propagation out to 1 AU.  This is equivalent to a change of -8.1$\mydeg$ in latitude and -26.4$\mydeg$ in longitude.  The majority of the deflection occurs while the CME is in the gradual rise and acceleration phases ($R\le$3$\rsun$).  By 5 $\rsun$ the CME comes close to a constant angular position: the CPA changes less than 1$\mydeg$ between 5 $\rsun$ and 1 AU. Fig. 6b shows that beyond about 10 $\rsun$ the CME's angular motion reverses and it slowly moves in the opposite direction.  This motion causes a change in the CPA of less than a degree and can be explained by a change in the direction of the forces acting upon the CME. 

The net deflection force comes from summing over the two CME edges.  Figure 7 shows the magnetic tension and magnetic pressure gradient force, respectively in red and blue, versus distance from the center of the Sun.  The figure also shows the total force (tension plus pressure gradient) in black.  Fig. 7 highlights the strongest forces, which occur close to the Sun.  Beyond 1.7 $\rsun$ the forces have decreased by several orders of magnitude from the values during the first few time-steps and are not included in the figure.   The force continues to decrease as the magnetic field decreases with distance.  

\begin{figure}
\epsscale{1.0}
\plotone{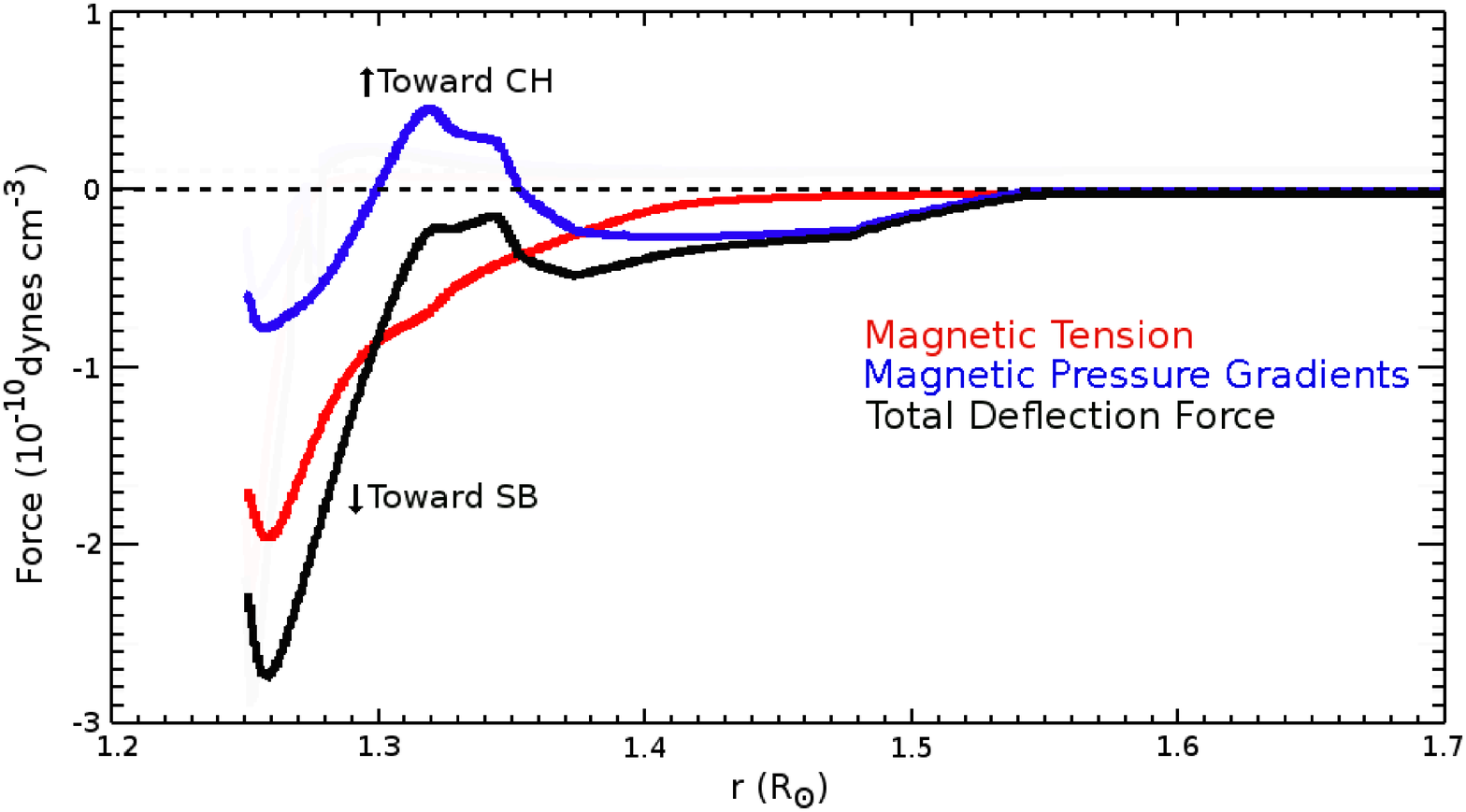}
\caption{The net forces acting on the CME.  The net magnetic tension is shown in red, net magnetic pressure gradients in blue, and the sum in black.}
\end{figure}

Initially both the magnetic tension and magnetic pressure gradients force the CME toward the SB, the tension force being about twice as strong as the magnetic pressure gradient force.  

The CME motion can be explained by considering the angular magnetic profiles in Fig. 5 as a series of potential barriers and wells since the deflection forces all depend on the magnetic field strength.  Initially, the magnetic pressure gradient force at the edge near the CH (hereafter CH edge) points toward the CH because of the strong magnetic field of the AR.  The magnetic pressure gradient force on the edge near the SB (hereafter SB edge) points towards the SB initially.  Because of the strong gradients near the SB, the magnetic pressure gradient force on the SB edge has a larger magnitude than the force on the CH edge so the net magnetic pressure gradient force points toward the SB.  The tension force always points toward the CME center for each edge so the direction of the total tension force will always be toward the edge in the weakest background magnetic field.  Initially, the net magnetic tension points towards the SB.  Both of these forces cause the CME to start moving toward the SB.

As the CME moves toward the SB the CH edge will interact with the potential barrier of the AR.  The CH edge in the control case starts close to the AR maximum so it quickly reaches the peak in $B(\phi )$ when the CME is at a distance of 1.26 $\rsun$.  After crossing the peak the magnetic pressure gradient force on the CH edge changes sign as the edge moves toward the PIL.  This force again changes direction as the CH edge crosses the PIL, then one final change at 1.36 $\rsun$ as it crosses the second maxima of the AR and continues the motion toward the SB.  Until 1.34 $\rsun$ the magnetic pressure gradient force on SB edge continues to point toward the SB.  The SB edge then crosses the minimum in $B(\phi )$ at the SB so the magnetic pressure gradient force switches direction.  The CME continues to move toward the SB until the SB edge pushes far enough into the SB potential well for the forces on the SB edge to overcome the forces on the CH edge.  After decelerating the SB-directed motion the magnetic pressure gradient forces cause the CME to begin move away from the SB.  By the time this occurs, the CME is several radii from the Sun so the force is minimal compared to the forces that initiated the deflection process.  However, this process does cause the CPA to change by a little less than a degree between 5 $\rsun$ and 1 AU.

The edge positions also affect the contribution of the magnetic tension force.  The tension force does not vary substantially as a result of the CH edge's motion through the PIL.  Until the CPA reaches the SB, the CH edge remains in higher background magnetic field strength than the SB edge so the tension force always pushes the CME toward the SB.  The tension force decreases quickly with both time and distance as the CME expands and moves away from the Sun toward regions of lower magnetic field strength.

CMEs deflected only as a result of magnetic forces will always head toward the minima in the magnetic field.  Observations have shown that CMEs do tend to head toward the HCS \citep{Kil09, Gui11, She11}.  For the control case, the magnetic background possesses strong global magnetic gradients.  These gradients cause the CME to reach the SB. For other Carrington Rotations with weaker global magnetic gradients this might not be the case.  The model is also limited by the inclusion of only magnetic deflection forces.  Other factors, not included in ForeCAT, such as interactions with other CMEs, effects of spatial variations in solar wind speed, or reconnection may still affect observed CMEs. ForeCAT does not include different plasma properties (density and temperature) for the SB compared to the rest of the solar wind background, which may also affect the SB's interaction with a CME.  Streamer blowouts should occur with a different background with weaker magnetic gradients.
  
\section{Parameter Sensitivity: Potential Deflection Angles}
We explore ForeCAT's sensitivity to the free parameters of the model, such as $M_{CME}$.  The mass is increased to 1x10$^{16}$ g while all other free parameters are unchanged.  Figure 8 shows the CME trajectory, analogous to Figure 6.
\begin{figure}
\epsscale{1.0}
\plotone{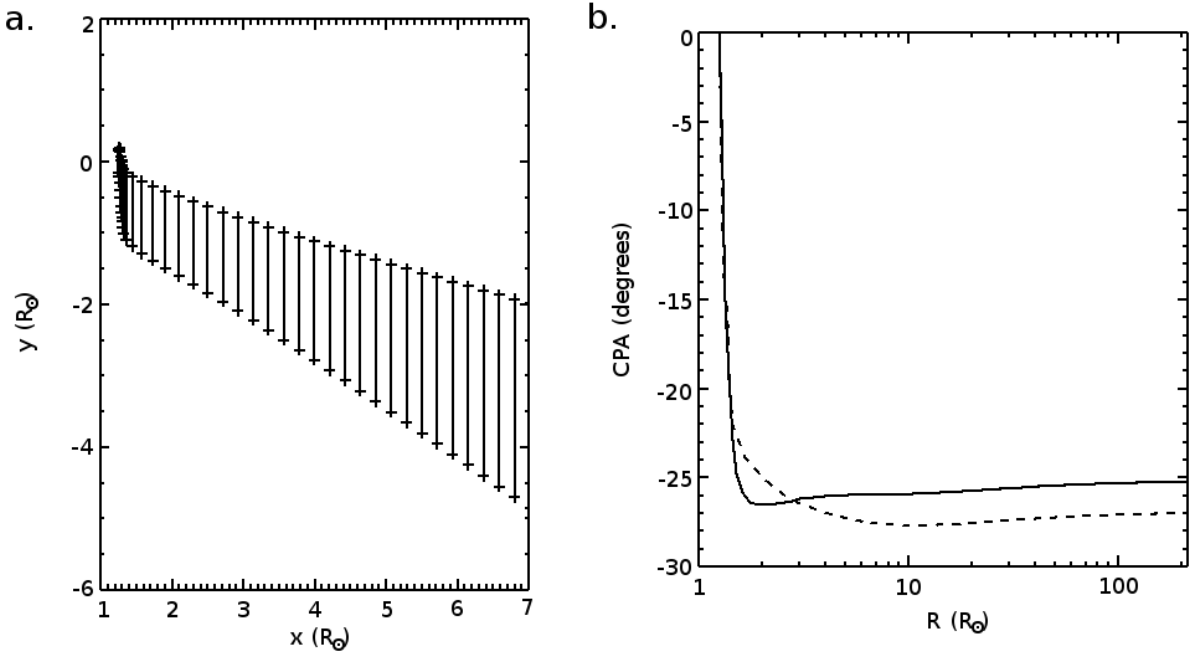}
\caption{Same as Figure 6 but for M$_{CME}$=1x10$^{16}$ g.  The amount of deflection decreases from -27.0$\mydeg$ in the control case to -25.3$\mydeg$ for the more massive CME.  The more massive CME deflects -7.6$\mydeg$ in latitude and -24.6$\mydeg$ in longitude.  Panel b shows the more massive CME with a solid line and includes the control case as a dashed line. The more massive CME pushes further into the SB close to then Sun causing a stronger force pushing it away from the SB, leading to a smaller final deflection.}
\end{figure}

\begin{figure}
\epsscale{1.0}
\plotone{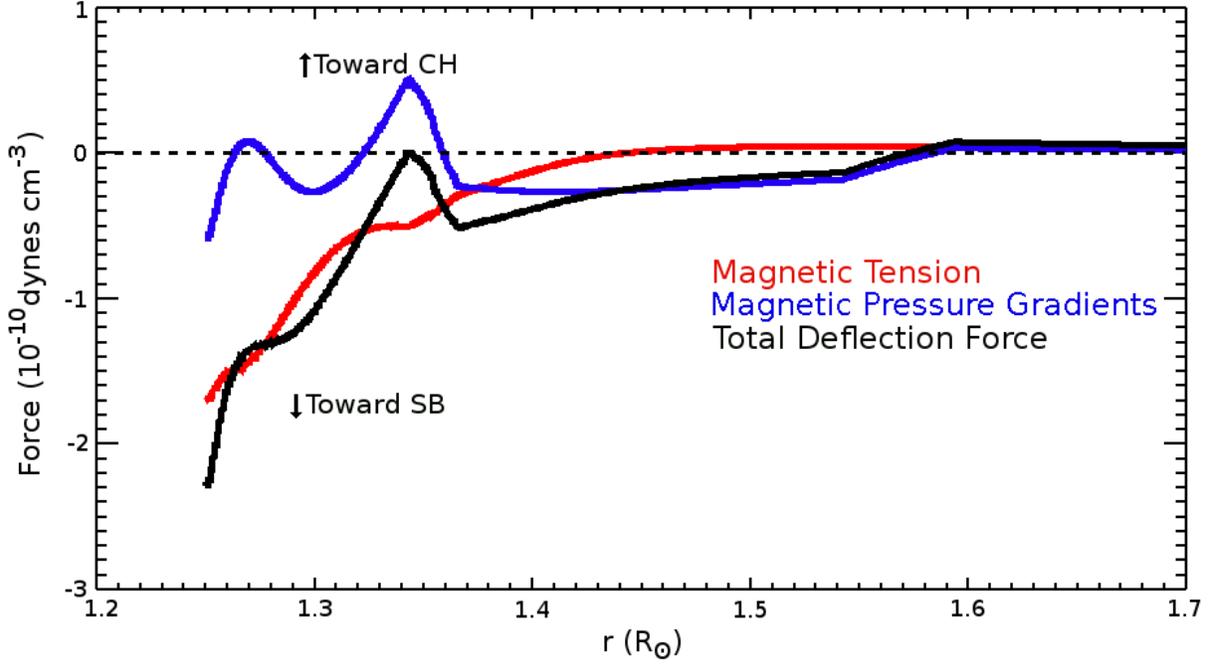}
\caption{Same as Figure 7 but for $M_{CME}$=1x10$^{16}$ g.  As with the control case, the forces initially push the CME toward the SB.  As the edges begin interacting with the AR and SB the magnetic pressure gradient force changes sign.  The more massive CME crosses the potential minimum closer to the Sun than the control case.  This leads to stronger forces in the opposite direction, causing a greater change in the CPA. The CME pushes far enough into the SB to produce a net magnetic tension force pointing toward the CH, as well as more noticeable positive magnetic pressure gradient force beyond 1.6 $\rsun$.}
\end{figure}

The more massive CME deflects -25.3$\mydeg$, 1.7$\mydeg$ less than the original case which deflected -27.0$\mydeg$.  At each distance, the deflection forces have comparable magnitudes to those on the control case.  Again these forces initially deflect the more massive CME toward the SB.  Figure 9 (analogous to Fig. 7 for the control case) shows that as with the control case, the forces change direction as the CME interacts with the AR and potential well of the SB.  The more massive CME crosses the SB minimum in B($\phi$) closer to the Sun so when the forces change sign they have a larger magnitude than when this occurs for the control case.  This causes the nonradial motion of the more massive CME to slow down faster than for the control case. The more massive CME also penetrates further into the SB potential well.  Around 1.45 $\rsun$ the magnetic tension changes directions when the background magnetic field strength near the SB edge becomes stronger than that near the CH edge.  Beyond 1.6 $\rsun$ the magnetic pressure gradient force is stronger than the control cases values. The strength and direction of these forces cause the CME to move further away from the SB, ultimately yielding a decrease in total deflection.

Next, deflection angles were calculated by varying both $M_{CME}$ and $v_f$, as shown in Figure 10a, and $B_0$ and $L_0$, as shown in Figure 10b.  The color indicates the change in the CPA between 1 $\rsun$ and 1 AU.
  
\begin{figure}
\epsscale{1.0}
\plotone{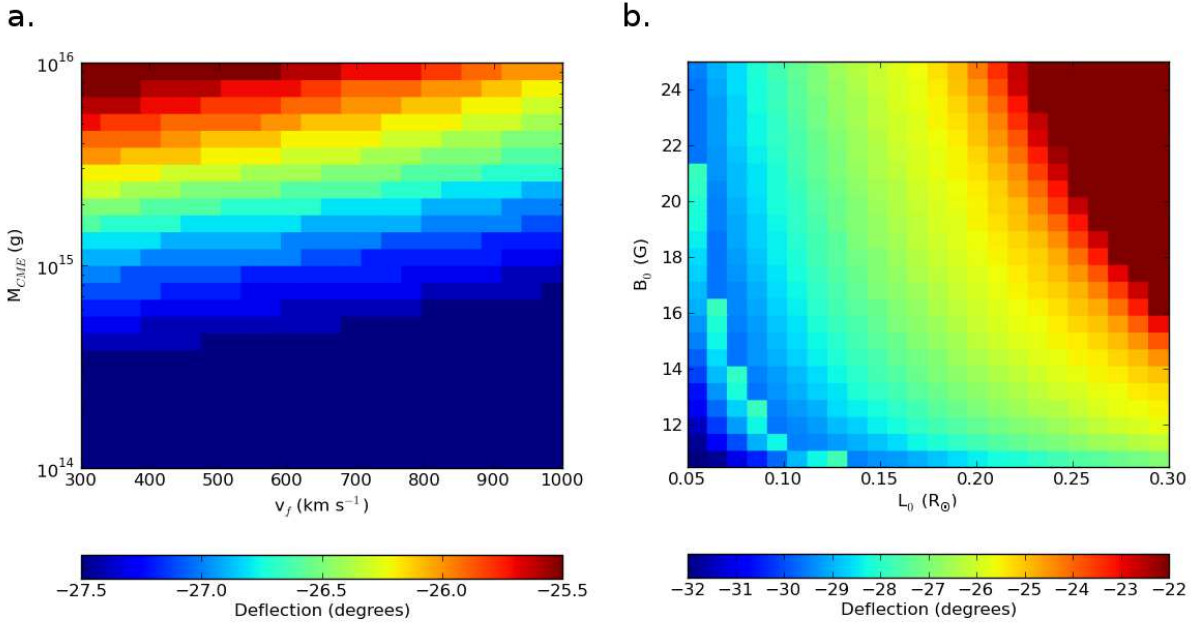}
\caption{Panel a shows contours of deflection versus the CME mass, $M_{CME}$, and final CME propagation speed, $v_f$.  Panel b shows deflection contours for the initial CME magnetic field strength, $B_0$, which drives the expansion, as well as for the initial CME radius, $L_0$.  The range for the calculated deflection angles in panel a is 2$\mydeg$.  For panel b the majority of the cases fall within a range of 10 $\mydeg$. A subset of CME with large initial size and large expansion fall outside this range and have deflections between -13$\mydeg$ and -16$\mydeg$.}
\end{figure}

Figure 10a shows the total deflection for varying CME mass between 10$^{14}$ and 10$^{16}$ g and varying the final propagation speed between 300 and 1000 km s$^{-1}$ while keeping all other parameters fixed.  The contour plots come from a sample of 625 CMEs (25$\times$25).  As seen in the individual test cases, more massive CMEs tend to deflect less.  The variation for masses ranging over two orders of magnitude is only 2$\mydeg$ showing that the strong gradients in this magnetic background force everything to the SB.  Faster CMEs tend to deflect less, but for any single mass the variance with velocity is half a degree.  Slower radial velocity causes a decrease in deflection similar to the behavior observed in the test cases.  The slower CME spends more time in a region with high forces causing it to reach the SB closer to the Sun.  As with the high mass CME the strong forces move the CME further back toward the direction it came from, decreasing the total deflection.

Figure 10b shows contours of the deflection versus the initial CME magnetic field, $B_0$, and initial CME cross-section radius, $L_0$.  These parameters determine the evolution of the CME's volume.  $B_{0}$ is varied from 10 to 25 G and $L_{0}$ between 0.05 and 0.30 $\rsun$.  Magnetic field strengths less than 10 G did not provide sufficient expansion to stop the CME from collapsing in on itself due to the large magnetic tension forces from the strong magnetic field background.  These parameters lead to three distinct populations with different deflection behavior.  The majority of the CMEs follow a pattern similar to that of the test cases.  Stronger expansion and larger initial size tend to lead to less deflection.  The second population is located in the lower left side of Fig. 10b below the band between (0.05 $\rsun$, 21 G) and (0.13 $\rsun$, 10G) corresponding to a deflection of -28$\mydeg$.  The smallest CMEs with weakest expansion had the strongest deflection and show slightly different behavior.  During propagation to 2 $\rsun$ these CMEs quickly deflect around 10-15$\mydeg$.  The CH edge of the CME does not cross the angular position of the AR in this time.  Between 2 $\rsun$ and about 60 $\rsun$ the CH edge of the CME moves toward the SB and crosses the first magnetic maximum (corresponding to the angular position of one polarity flux system of the AR) causing an additional 5$\mydeg$ of deflection.  Near 60 $\rsun$ the CH edge passes over the final magnetic maximum (corresponding to the angular position of the other polarity flux system of the AR) leading to an additional 10-15$\mydeg$ of deflection by 1 AU.  Of the 625 CMEs,  41 CMEs display this sort of behavior.  The third population falls in the solid red region in the upper right corner of Fig. 10b.  The CMEs with the strongest expansion and initially large size tend to deflect substantially less than the other two sets.  90 of the CMEs deflect between -13 and -16$\mydeg$.  These CMEs deflect less because the CME reaches an equilibrium position with the SB edge near the SB minimum and the CH edge near the CH minimum.  Due to initial size, strong expansion, or some combination thereof, the CH edge of the CME never crosses over the magnetic maximum at the angular position of the AR.

The same method used to explore the influence of the CME properties can be used to analyze some of the parameters in the propagation model.  The assumed CME radial velocity profile may affect the net deflection.  For the control case the propagation parameters $R_{ga}$, $R_{ap}$, and $v_g$ were varied while all other parameters were unchanged.  The ranges for each parameter are 1.25 $\rsun\le R_{ga}\le$2.25 $\rsun$,  2.5 $\rsun \le R_{ap}\le$4 $\rsun$, and 25 km s$^{-1} \le v_g \le$100 km s$^{-1}$.  For these ranges, the final deflection angle varied by less than 0.2$\mydeg$.  Therefore we conclude that the parameters chosen for the propagation model do not greatly influence the deflection of the CME.

We explore the sensitivity of the deflection to different values of $C_d$.  Using MHD simulations \citet{Car96} show that values of $C_d$ between 1 and 3 are appropriate for the acceleration phase of a CME.  \citet{For06} use $C_d=\tanh(\beta)$ where $\beta$ equals the ratio of the thermal and magnetic pressure.  Close to the Sun, $\beta <<1$ so $C_d$ will be small. We use larger values than $\tanh(\beta)$, using constant values of $C_d$ between 0.25 and 10, similar to the range of $C_d$ in \citet{Car04}.  We find that these values yield deflections varying by 2$\mydeg$ for the control case.  Stronger drag causes less deflection but ultimately the CME still deflects to the SB because of the strong magnetic gradients specific to this background.  The drag changes the distance at which the CME begins interacting with the SB.  With a weaker background the chosen drag coefficient may have a more significant effect.  We explore as well other expressions of $C_d$ contained in the literature.  \citet{Sis06} contains two models of the drag coefficient versus distance: a linear model:
\begin{equation}
C_d=1+\frac{5R}{1 \; AU}
\end{equation}
and a quadratic model:
\begin{equation}
C_d=\left(1+\frac{1.45R}{1 \; AU}\right)^2
\end{equation}
These models produce deflections less than 0.01$\mydeg$ smaller than the control case with $C_d$=1.

\section{Deflection with PFSS Background}
We run the control case using a PFSS background to see the effects of the position of the SB varying with height.  The PFSS model uses the same set of coefficients as used to initialize the MHD solution.  As seen in Fig. 5b at 2.5 $\rsun$ the HCS forms around -5$\mydeg$, over 20$\mydeg$ away from the minimum in the scaled background.  Figure 11 shows the CPA versus distance for the control case with the PFSS background (in black) and the scaled background (in blue).

\begin{figure}
\epsscale{1.0}
\plotone{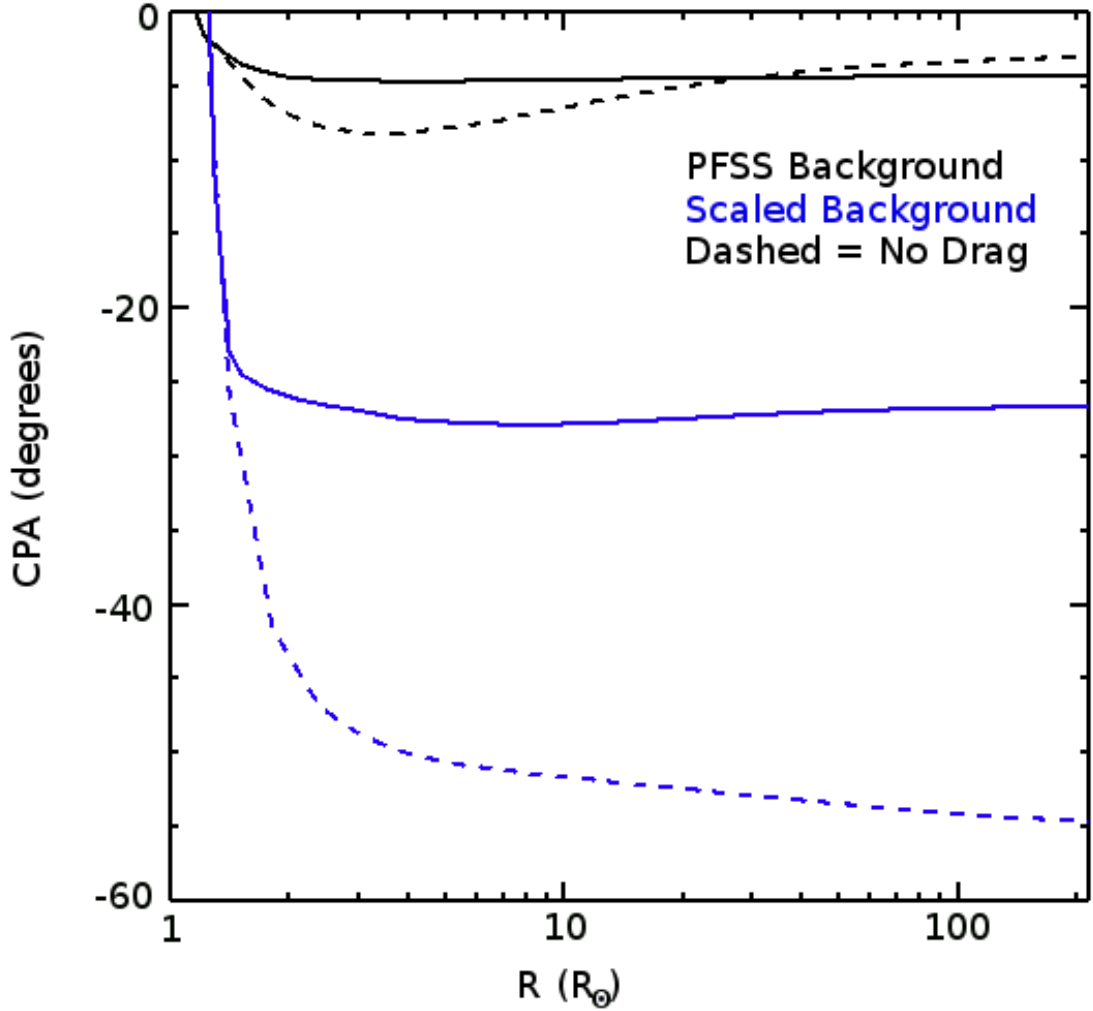}
\caption{Central position angle versus distance for the control case run with two different magnetic backgrounds.  The black line shows the results of a PFSS background, the blue line the scaled background.  In the PFSS background the minimum in the magnetic field moves to approximately -5$\mydeg$ and accordingly the PFSS CME only shows a deflection of -4.4$\mydeg$ (-1.6$\mydeg$ in latitude and -4.1$\mydeg$ in longitude) compared to the -27.0$\mydeg$ of the scaled background CME.  The dashed lines show each background run without including the effects of non-radial drag.}
\end{figure}

The PFSS control case deflects -4.4$\mydeg$, a deflection of -1.6$\mydeg$ in latitude and -4.1$\mydeg$ in longitude.  As seen in the scaled case, the CME deflects to the minimum in the magnetic field strength but that position has changed because of the nature of the PFSS model.  The forces of the PFSS model decrease much quicker with distance than the scaled model, however, with this case the magnetic minimum moves closer to the initial CME position.  For cases where the magnetic minimum is further from the initial CME position the rapid decrease in forces could cause the CME to only deflect in the direction of the minimum, not fully to it. In future work we will continue to explore the sensitivity of the deflection to the magnetic background.  To better understand the difference between the models requires comparisons within a background with weaker gradients or where the magnetic minimum moves away from the initial CME location.

Future work will also include the effects of including the nonradial drag force.  As mentioned in section 6, the deflection is sensitive to the chosen drag coefficient but complete exclusion of nonradial drag produces even more variation.  Fig. 11 includes both PFSS and scaled runs without drag as dashed lines.  The effect for the PFSS case is smaller since the total deflection is smaller but for the scaled model we see a difference of nearly 30$\mydeg$ in the cases with and without nonradial drag. 

\section{Effects of Active Regions}
We explore here the effects that an AR can have on a CME's deflection.  We define a new deflection plane based on the orientation of the AR.  Close to the Sun the magnetic pressure gradients exhibit complex behavior (Fig. 1a) and cannot be used to define the deflection plane.  Deflection from an AR will result from imbalances between the different polarity flux systems of the AR.  We define the overall gradient of the AR using the positions of the point within each polarity containing the strongest magnetic field.  This gradient replaces the gradient vector in the deflection plane calculation.  Figure 12 contains the scaled magnetic profile within the deflection plane calculated using this AR vector, analogous to Figure 5. 

\begin{figure}
\epsscale{1.0}
\plotone{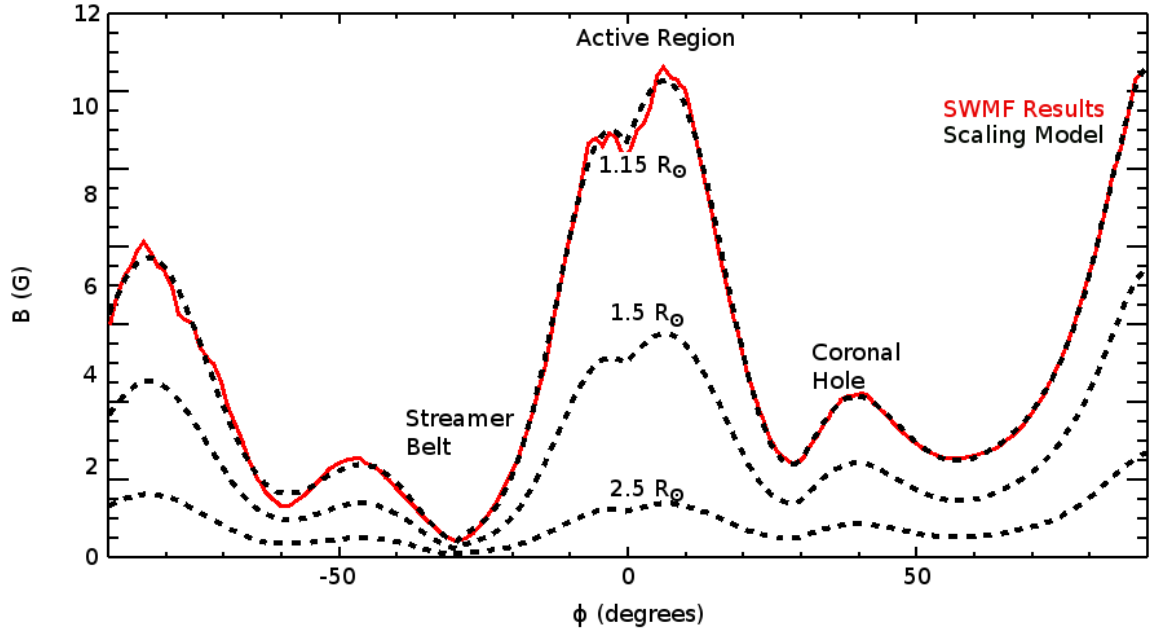}
\caption{Same as Figure 5, but for the magnetic field within the AR deflection plane.  The magnetic field has a local minimum at the polarity inversion line in between the two contributing flux systems of the AR.  The system closest to the CH has a stronger magnetic field than the system near the SB within this plane.}
\end{figure}

 The MHD model does not capture the full complexity of the magnetic field in an AR but it does include some variation between the opposite polarity flux systems, which is more pronounced in this plane than the original deflection plane.  The system near the CH has a stronger magnetic field than the system near the SB, and the magnetic field decreases near the polarity inversion line (PIL) between the two systems. 

The results presented here use the control case parameters (see Section 5), but launched from $\phi$=0$\mydeg$ within the new deflection plane, close to the local minimum corresponding to the PIL.  This CME is deflected -24.6$\mydeg$ during propagation out to 1 AU.  The global magnetic gradients still contribute within this plane and the heightened asymmetry between the opposite polarity flux systems drives additional deflection.  As a result, we determine that close to the Sun it may be necessary to redefine the deflection plane along the CMEs trajectory if we wish to accurately predict an actual CME's deflection.

\section{Discussion}
ForeCAT shows that magnetic forces alone can be responsible for significant CME deflection.  The model excludes several other factors known to deflect CMEs, such as interaction with other CMEs, spatially varying background solar wind, or reconnection, yet still results in deflections of comparable magnitude to observed deflections.  Calculating the magnetic forces along the CME trajectory relies on many fundamental assumptions about the CMEs behavior.  These assumptions may prevent the current version of ForeCAT from predicting the precise behavior of actual CMEs, but allow us to demonstrate the importance of magnetic forces for CME deflection.  Future works will continue to improve the model and refine some of the underlying assumptions.

The description of the CME flux rope used assumes that the toroidal axis of the CME lies within the deflection plane.  Any deviation from the CME being perfectly perpendicular to the deflection plane will result in an elliptical CME cross-section.  The CME cross-section contributes to the tension calculation.  To first order, a tilt between the toroidal CME axis and the normal to the deflection plane could be accounted for by calculating the radius of curvature as $\kappa = 1 / (L \sin(\beta))$ where $\beta$ is the angle between the deflection plane and the toroidal CME axis. The present version of ForeCAT assumes that $\beta$=90$\mydeg$ which will lead to an underestimate of the curvature and the magnetic tension force for an elliptical CME cross-section.

CME-driven shocks would also distort the draping of the background magnetic field around the CME.  Shocks are known to change the orientation of the magnetic field: fast-mode shocks rotate the magnetic field away from the shock normal, slow-mode shocks rotate the magnetic field toward the shock normal.  A shock would cause the background magnetic field to rotate which would affect how the field then drapes around the CME.  The draping out of the deflection plane would affect not only the direction of the magnetic tension force, but the magnitude may change as well if the magnetic field drapes around a region of the CME with different curvature. 

The calculation of the forces is restricted to the two points where the deflection forces should be the strongest.  In reality the force should be integrated over the full surface of the CME. By assuming a solid CME body we assume as well that the motion of the CME cross-section in the deflection plane applies to the entire CME.  Different cross-sections will feel different forces which should be accounted for when trying to compare to specific observations. This will be addressed in a future work.

The assumption of a solid CME body also affects the ForeCAT results.  Close to the Sun deflection, rotation, and nonuniform expansion cannot be distinguished.  \citet{Nie12} reconstruct the 3D trajectory of the 16 June 2010 CME, a CME with significant rotation.  They find that not accounting for the CME rotation can cause substantial errors in the calculation of the CME size leading to an overestimate of the latitudinal CME expansion.  Using a global MHD model, including deflection, rotation, and expansion, \citet{Eva11} present three simulations of flux ropes with different orientations embedded in the same background solar wind.  They show that the resulting shape and dynamical evolution of the CME are highly dependent on the initial CME orientation.  In general, external forces likely will cause a combination of deflection, rotation and deformation.  What is interpreted in coronagraph observations as rotation or nonuniform expansion could be differential deflection along the CME.  We assume that the external forces cause the CME to deflect rather than to deform.  The rotation and non-uniform expansion will be addressed in a future study.

The magnetic background is assumed to remain in a static configuration.  However reconnection can occur during the eruption and evolution of the CME. The reconnection will alter the background magnetic field, transforming magnetic energy into kinetic and thermal energy, potentially affecting the background magnetic pressure gradients. In addition, when  magnetic field lines draping around the CME reconnect the tension force will become unbalanced leading to deflection similar to that of \citet{Lug11} and \citet{Zuc12}.

The CME's radial motion is predetermined using the analytic model.  A CME's motion results, at least in part, from the same Lorentz force that drives the magnetic deflection.  As the CME deflects to regions of weaker magnetic background the radial acceleration of the CME will be affected as well.  Feedback between the radial and non-radial motion could lead to deflections different from those determined with an analytic radial propagation model.

\section{Conclusions}
This manuscript presents ForeCAT, a model of CME deflection using deflection forces from both magnetic pressure gradients and magnetic tension. ForeCAT relies on many simplifying assumptions, but several test cases show that magnetic forces alone can cause deflections of similar magnitude as observed deflections.   Future work will refine these assumptions and allow for comparisons between ForeCAT results and specific observed CMEs.  The current ForeCAT model has already yielded several insights into CME deflection.

The magnetic forces cause the CME to deflect towards the SB, the minimum in the magnetic field strength.  For most CMEs the magnetic forces are sufficiently strong that the majority of the deflection occurs within the first several solar radii.  The chosen magnetic background contains strong magnetic gradients so that the deflected CME reaches the SB.  In a weaker background, deflection will move CMEs toward the SB, but may not be capable of fully deflecting the CME to the SB.  Deflection will also change due to interactions with other CMEs, spatially varying background solar wind velocities, or reconnection, the effects of which are not included in ForeCAT. The inclusion of variations in temperature and density in the background plasma could also affect how the CME reaches its equilibrium angular position.

An exploration in parameter space shows variation in the final deflection for a wide range of input parameters, for this specific background.  The majority of CMEs deflect fully to the SB within a couple solar radii.  Two different subsets of CMEs exhibit different behavior.  Initially small CMEs with little expansion deflect the most.  These CMEs do not reach the SB close to the Sun but instead partially deflect and continue along a nearly constant angular position until a secondary period of strong deflection occurs around 60 $\rsun$.  This secondary deflection deflects the CMEs further since they interact with the SB much further from the Sun where the gradients are weaker.  Initially large CMEs with strong expansion deflect the least.  These CMEs remain above the AR, an equilibrium angular position is found with each edge in a potential minimum on either side of the AR.  The relative strength of the magnetic minima at the SB and the CH cause a slight deflection toward the SB.  

The PFSS background yields different deflections than the scaled background.  The scaled background contains a more realistic radial dependence assumes that the angular dependence is fixed.  The PFSS background decreases too quickly with radial distance, but the angular location of the streamer belt is not fixed.  These two differences between the models lead to differences in deflection, 4.4$\mydeg$ compared to 27$\mydeg$ in the PFSS and scaled backgrounds, respectively.  CME deflections depend strongly on the magnetic background which will be a focus of future work.

ForeCAT can be extended to uses other than just solar CMEs.  Using the AR deflection plane and a more complex model of the AR magnetic field, ForeCAT should be able to capture the rolling motion of prominences.  Given some approximation of the background magnetic field, ForeCAT can probe the space weather conditions of planetary systems around other stars.  The magnetic fields of low mass stars can reach several kG \citep{Rei07}, far exceeding solar values, so significant CME deflections could occur depending on the properties of the ejecta.

\acknowledgments
The authors would like to thank the anonymous referee for his/her comments.  C. K. is supported by NSF CAREER ATM-0747654.  R. M. E. is supported through an appointment to the NASA Postdoctoral Program at GSFC, administered by Oak Ridge Associated Universities through a contract with NASA.


\appendix

\section{Overpressure Expansion}
The equation for expansion results from a series of modifications to a classic hydrodynamics problem of a spherical over pressured region expanding in a fluid \citep{Mil68}.  This appendix presents the derivation of the final equation, starting from the hydrodynamics.  The momentum equation can be written as
\begin{equation}
\frac{d\vec{v}}{dt} = \vec{F} - \frac{1}{\rho}\nabla P
\end{equation}
where $\vec{v}$ is the velocity, $\vec{F}$ represents external forces, $\rho$ is density and P is pressure.  The total velocity derivative has two contributions: the local and convective components.  For incompressible fluids, the convective term becomes $\frac{1}{2}\nabla v ^2$.  Rearranging gives
\begin{equation}
\frac{\partial \vec{v}}{\partial t} = - \nabla \left( \int\frac{\partial P}{\rho}+\frac{1}{2}v^2\right)
\end{equation}
where no external forces is assumed.  For an irrotational fluid the velocity can be written as the negative gradient of a scalar field ($v=-\nabla\phi$) so that A2 becomes
\begin{equation}
\nabla\left( \int\frac{\partial P}{\rho}+\frac{1}{2}v^2 - \frac{\partial \phi}{\partial t}\right)=0
\end{equation}
or integrating
\begin{equation}
\frac{P}{\rho}+\frac{1}{2}v^2 - \frac{\partial \phi}{\partial t} = C(t)
\end{equation}
The formalism of a fluid source can be used to simplify equation A4.  A source emits $4\pi m$ of volume per time, where $m$ is the strength of the source.  Applying conservation of mass in 3D and assuming only radial velocities ($v=v_r$) gives $m=r^2v$.  Plugging this into $v=-\nabla\phi$ and integrating both sides with respect to $r$ yields $m=\phi r$ or $v=\frac{\phi}{r}$.  
\begin{equation}
\frac{P}{\rho}+\frac{1}{2}\left(\frac{\phi}{r}\right)^2 - \frac{\partial \phi}{\partial t} = C(t)
\end{equation}
The spherical overexpanding cavity is considered using the fluid source description.  Assuming at some time the cavity has radius $R$, at the edge of the cavity ($r=R$) the change in radius is defined to be $R'$ (the same as $v$ since the velocity is only radial) which corresponds to $\phi=RR'$ giving a source strength $m=R^2R'$.  The scalar field then has an $r$-dependence
\begin{equation}
\phi = \frac{R^2R'}{r}
\end{equation}
where $r$ is not just limited to the radius of the cavity.  Taking the partial time derivative of A6 and rewriting A5 gives 
\begin{equation}
\frac{P}{\rho}+\frac{1}{2}\left(\frac{R^2R'}{r^2}\right)^2-\frac{R^2R''+2RR'^2}{r}=0
\end{equation}
where $C$ is set equal to zero because as $r$ goes to infinity the pressure should be negligible.  Looking at $r=R$ 
\begin{equation}
\frac{P}{\rho}+\frac{1}{2}(R')^2-RR''+2R'^2=0
\end{equation}
or
\begin{equation}
R''=\frac{1}{R}\left(\frac{P}{\rho}-\frac{3}{2}(R')^2\right)
\end{equation}
For a cavity dominated by the magnetic pressure $P \propto B^2$.  Assuming mainly poloidal magnetic field then $B$ must fall as $R^{-2}$ to conserve magnetic flux.  The pressure then changes as
\begin{equation}
\frac{P}{P_0}=\left(\frac{R_0}{R}\right)^{4}
\end{equation}
and assuming  a magnetic overpressure
\begin{equation}
P=\frac{B^2_{CME}}{8 \pi}-\frac{B^2_{SW}}{8 \pi}
\end{equation} 
Dividing by the initial solar wind density $\rho_{SW0}$ gives
\begin{equation}
P_0=\rho_{SW0}\frac{A^2_{h0}-A^2_{SW0}}{2}
\end{equation}
where $A_{h0}$ is a hybrid Alfv\'en speed using the CME initial overpressure strength and the initial background solar wind density whereas $A_{SW0}$ uses the initial solar wind magnetic field.  Using A9, A10, and A12 leads to a final expression
\begin{equation}
R''=\frac{1}{R}\left(-\frac{3}{2}(R')^2 +\frac{\rho_{SW0}}{\rho}\left(\frac{R_0}{R}\right)^4\left(\frac{A^2_{h0}-A^2_{SW0}}{2}\right)\right)
\end{equation}
which is the same as equation 12 with $L$ replacing $R$.

\clearpage

\end{document}